\documentclass[mathleft
]{an}
\usepackage{graphicx}
\usepackage{times}
\usepackage{natbib}
\bibpunct{(}{)}{;}{a}{}{}
\begin{document}
% The following seven commands are intended for editorial usage and should be ignored by
% the author(s).
\Pagespan{1}{}% Document's page range.
% If second parameter is left empty, the last page is computed automatically.
\Yearpublication{2006}%
\Yearsubmission{2005}%
\Month{11}%
\Volume{999}%
\Issue{88}%
% \DOI{This.is/not.aDOI}%

\title{Modeling the total and polarized emission in evolving galaxies:
``spotty'' magnetic structures}

\author{T.G. Arshakian,\inst{1}\fnmsep\thanks{On leave from the Byurakan
Astrophysical Observatory, Aragatsotn prov. 378433, Armenia, and
Isaac Newton Institute of Chile, Armenian Branch \newline}
R. Stepanov,\inst{2} R. Beck,\inst{1} M.
Krause,\inst{1} and D. Sokoloff\inst{3}
}
\titlerunning{Modeling the magnetic fields}
\authorrunning{T.G. Arshakian et al.}
\institute{
MPI f\"ur Radioastronomie, Auf dem H\"ugel 69, 53121
Bonn, Germany \\
 \email{[tarshakian;rbeck;mkrause]@mpifr-bonn.mpg.de}
\and
Institute of Continuous Media Mechanics, Korolyov
str. 1, 614013 Perm, Russia, \\ \email{rodion@icmm.ru}
\and
Department of Physics, Moscow State University, Russia \\ \email{sokoloff@dds.srcc.msu.su}
}

\received{30 Jan 2010}
\accepted{11 Feb 2010}
\publonline{later}

\abstract {Future radio observations with the Square Kilometre Array
(SKA) and its precursors will be sensitive to trace spiral galaxies
and their magnetic field configurations up to redshift $z\approx3$. We
suggest an evolutionary model for the magnetic configuration in
star-forming disk galaxies and simulate the magnetic field
distribution, the total and polarized synchrotron emission, and the
Faraday rotation measures for disk galaxies at $z\la 3$. 
%Under consideration of the structure of the seed field and the star-formation
%rate and the influence of encounters and mergers during the galaxy
%evolution. 
Since details of dynamo action in young galaxies are
quite uncertain, we model the dynamo action heuristically relying
only on well-established ideas of the form and evolution of magnetic
fields produced by the mean-field dynamo in a thin disk. We assume a
small-scale seed field which is then amplified by the small-scale
turbulent dynamo up to energy equipartition with kinetic energy of
turbulence. The large-scale galactic dynamo starts from seed fields
of 100 pc and an averaged regular field strength of 0.02\,$\mu$G,
which then evolves to a ``spotty'' magnetic field configuration in
about 0.8\,Gyr with scales of about one kpc and an averaged regular
field strength of 0.6\,$\mu$G. The evolution of these magnetic spots
is simulated under the influence of star formation, dynamo action,
stretching by differential rotation of the disk, and turbulent
diffusion. The evolution of the regular magnetic field in a disk of
a spiral galaxy, as well as the expected total intensity, linear
polarization and Faraday rotation  are simulated in the rest frame
of a galaxy at 5\,GHz and 150\,MHz and in the rest frame of the
observer at 150\,MHz. We present the corresponding maps for several
epochs after disk formation. Dynamo theory predicts the generation
of large-scale coherent field patterns (``modes''). The timescale of
this process is comparable to that of the galaxy age. Many galaxies
are expected not to host fully coherent fields at the present epoch,
especially those which suffered from major mergers or interactions
with other galaxies. A comparison of our predictions with existing
observations of spiral galaxies is given and discussed.}


\maketitle
%
%________________________________________________________________

\section{Introduction}

Magnetic fields of nearby galaxies demonstrate several magnetic
configurations and related configurations of polarized radio
emission including a ``ring'' of polarized emission in M\,31, M\,33
\citep{tabatabaei08} and magnetic arms in NGC\,6946 \citep{beck05a}.
A particular magnetic configuration in a galaxy is believed to be
determined by specific features of the excitation mechanism. In
other words, magnetic configurations trace important details of
galactic MHD. For example, the magnetic configurations of the barred
galaxies NGC\,1097 and NGC\,1365 trace gas flows presumably feeding
the central active galactic nuclei \citep{beck05b}.

Future radio observations with the Square Kilometre Array (SKA) are
expected to give rich information concerning magnetic configurations
in the disk of the first galaxies with redshifts up to at least
$z=3$ \citep{arshakian09}. The aim of this paper is to suggest a
plausible evolutionary model for the magnetic configuration in early
galaxies and to simulate the magnetic field distribution,
and polarized synchrotron emission, and the Faraday rotation
measures for disk galaxies at $z\la 3$ (the polarization survey in
the frame of the SKA Design Studies) which can be tested by future
observations with the SKA and its precursors.

On first sight, the answer on the question under discussion looks
quite straightforward. The mean-field dynamo is an amplification and
ordering mechanism which excites one or few modes of regular
magnetic fields. The mode with maximal growth rate should dominate
very soon, independent of the configuration of the seed magnetic
field, and determine together with the distribution of relativistic
and thermal electrons the pattern of polarized emission. Later, the
magnetic field becomes dynamically important and the dynamo becomes
nonlinear, while the main shape of the configuration survives.

A closer look, however, shows that it is much more complex. The
first point is that galactic disks are thin and the time required
for a dynamo to establish the magnetic field pattern perpendicular
to the galactic plane, $t_w$, is much shorter than the corresponding
time required to elaborate the configuration in radial extent of the
disk, $t_r$ \citep{arshakian09}. (We use a cylindric coordinate
system $r, \phi, w$ connected with the equatorial plane of the
disk.) In particular, the diffusion time scales with distance as
$\sim r^2$. If we assume that $t_r$ and $t_w$ are proportional to
the corresponding diffusion times, then $t_r/t_w = (R/h)^2 \approx 2
\times 10^{3}$ for a disk aspect ratio $h/R \approx 2\times 10^{-2}$
(here $h$ and $R$ are the vertical and radial scalelengths of the
ionized gas).

The second point is that if the seed
field was not ordered at the overall galactic scale, the growing
regular magnetic field can preserve alternating directions in large
regions for a long time. As suggested by \cite{poezd93} \citep[see
also][]{beck94}, a suitable seed magnetic field in galactic disks
can be produced by the fluctuation dynamo. It is then natural to
expect that a large-scale magnetic structure with reversals along
radius \citep{ruzmaikin85} and/or azimuth \citep{bykov97} can
persist for a period comparable to the galactic lifetime.

The third point is that $t_w$ has to be shorter than the galactic
lifetime. However, $t_r$ may be comparable to the galactic age and
the magnetic field becomes dynamically important before the dynamo
develops a unique magnetic configuration over the whole galaxy. Then
the spotty magnetic field structure evolves nonlinearly and can
remain complicated for a rather long time. Observations of magnetic
fields in nearby galaxies confirm that their magnetic configurations
are more complicated than just a simple leading dynamo eigenmode
\citep{beck05a}. 

In this paper, we try to address these points in a simplified way.
Summarizing, observations of magnetic configurations and their
evolution with redshift $z$ can be a tracer for the structure of the
seed magnetic field and its origin. Obviously, events like
encounters, mergers, etc. make this picture much more rich and also
more complicated but this is beyond the scope of this paper. 

In Sect.~\ref{ssec:phys_bases}, we present the physical bases of the dynamo-evolution model.
The model of the evolution of magnetic
fields in disk galaxies is described in Sect.~\ref{sec:model}. The link between star-formation rate and total magnetic field of a galaxy is presented in Sect.~\ref{sec:link}, and modeling of the regular magnetic fields,
total intensity, polarization and Faraday rotations of evolving
galaxies are presented in Sect.~\ref{sec:simul}. Discussion and conclusions are
given in Sect.~\ref{sec:discussion}.

%\section{Model of evolving magnetic fields in galaxies}

\section{Physical picture of evolving regular magnetic fields}
\label{ssec:phys_bases}

A three-phase model of the evolution of regular magnetic fields in
galaxies was developed by \cite{arshakian09}. The dynamo theory,
which successfully describes the polarized synchrotron emission and
Faraday rotation in nearby galaxies, was used to derive the
timescales of amplification and ordering of magnetic fields on small
and large scales in disk and quasi-spherical galaxies. This allows
predicting the generation and evolution of magnetic fields in young
galaxies at high redshifts. We refer to disk galaxies that rotate
differentially and the turbulence is driven by supernovae, 
and we adopted the standard values of turbulence
velocity $v=10$ km s$^{-1}$ and a basic scale of turbulence of $l=100$
parsecs.
%Turbulence in galaxy disks is assumed to be driven by
%supernovae \citep{ruzmaikin88, korpi99,avillez05,gressel08}. 
%The other basic parameters of the galactic mean-field dynamo are the
%angular velocity of the galactic rotation $\Omega$, the disk
%thickness $h$, the disk radius $R$, and the gas density $\rho$.
Results from simulations of hierarchical structure formation
cosmology provide a tool to develop an evolutionary model of regular
magnetic fields coupled with galaxy formation and evolution.

In the first phase ($z\sim20$), the seed magnetic fields of order 
$\simeq 10^{-18}$~G \citep[][and references therein]{arshakian09} were generated in dark
matter galactic halos by the \emph{Biermann battery mechanism} or
the \emph{Weibel instability} (Weibel 1957; Schlickeiser \& Shukla 2003; Lazar et al. 2009).
Weak seed magnetic fields can be exponentially amplified by the
small-scale dynamo during the formation of the first stars
\citep{schleicher10,sur10} and by the Biermann mechanism in
supernova explosions of the first stars \citep{hanayama05}.

The key feature of this model is that, in the second phase, the
accreted gas from the intergalactic medium and mergers of small-mass
dark-matter halos generated turbulence on scales comparable with the
size of the infalling gas in the halo of a protogalaxy. The
small-scale dynamo driven by turbulent, shock-heated gas in the halo
of a protogalaxy can rapidly amplify the seed field from $\sim
10^{-18}$\,G to the equilibrium level, $\sim 10^{-5}$\,G, in a
relatively short time, $\sim 4\times 10^{8}$ years. This phase was
for the first time proposed by \cite{arshakian09} and it may happen
at $z\sim10-20$ or even earlier.

In the third phase, \emph{the extended thin disk} was formed 
at redshifts from $z\la 10$. The formation of gas-rich disks can be
triggered by the dissipation of the protogalactic halo or by major
merger events. A weak, large-scale magnetic-field component was
generated in the disk from small-scale magnetic fields of the halo
(amplified to the equilibrium level in the second phase) and further
amplified to the equipartition level by the mean-field dynamo within
few billion years. It was assumed that the small-scale galactic dynamo 
is acting together with the large-scale dynamo and that it  
saturates at the equipartition with the kinetic energy of interstellar turbulence \citep[e.g.][]{subramanian98}.
The evolution of the coherence scale of the
regular large-scale magnetic fields took a longer time: full
coherence was reached in dwarf galaxies within $\sim 6$\,Gyr and in
disk galaxies within $\sim 8$\,Gyr. Giant galaxies need more than
15\,Gyr to develop a fully coherent field.

The above picture is not the only possibility. The seed magnetic
fields could as well be produced by the turbulent dynamo in an
already formed galaxy as soon as turbulence driven by star formation
develops (more precisely, in $10^7$--$10^8$ years after the first
star formation burst). Then the first two stages are unimportant,
but the form of the seed magnetic field for the large-scale dynamo
action remains the same as in the above picture (but delayed). Therefore,
independent of the poorly known details of galactic formation and
their early evolution, we can more or less confidently suggest a
plausible and robust distribution of the seed magnetic field (see
Sect.~\ref{spots}).

The dynamo operates in the disk and is able to amplify the amplitude
of the regular field and to order the field on timescales given in
\cite{arshakian09}.

\section{Model of evolving magnetic fields in galaxies}
\label{sec:model}

\subsection{Origin of seed magnetic fields in first galaxies: magnetic spots}
\label{spots}

The large-scale galactic dynamo starts from a configuration
disordered on the scales of the disk radius and inevitably gives a
``spotty'' distribution of magnetic fields over the galactic disk.
Magnetic fields in nearby spots have opposite directions (clockwise
and counterclockwise). It looks plausible that the spotty nature of
the magnetic field (see Fig.~3) can be isolated observationally,
provided the resolution and sensitivity allow observations in
galaxies with suitable redshifts.

There is however an alternative viewpoint. The early Universe
could contain magnetic fields. This field could be
homogeneous and strong enough to serve as a seed field for the
galactic dynamo, while weak enough to keep the Universe practically
homogeneous and isotropic. Then the seed magnetic field is coherent
on the scales much larger than the galactic size, and one can hope
in principle to avoid a phase with ``spotty'' magnetic structure and
obtain a magnetic field ordered on the scale of the whole galaxy at
all stages of galactic evolution. This scenario looks less
attractive, but it cannot be rejected by the available observations
and theoretical ideas. Moreover, the existence of a homogeneous
cosmological magnetic field gives an important message for
cosmology. Causal mechanisms for magnetic field generation in the
early Universe, based on standard practical physics, result in
magnetic fields with contemporary scales which are negligible
compared to galactic scales. This means that the desired primordial
homogeneous magnetic field wo\-uld have an a-causal nature, i.e. the
scale was larger than the horizon at the time of its creation.

It looks potentially possible to distinguish between the above
scenarios by future observations with the SKA and hence support or
reject the option of an a-causal primordial magnetic field. In case
of the observationally confirmation of the a-causal idea, primordial
magnetic field will be supported observationally, the concept of
standard particle physics have to be reconsidered
\citep[e.g.][]{giovannini04,semikoz05}.

Obviously, both scenarios require to be elaborated. To be specific,
we present here the results concerning the first, most probable
scenario while the evolution of the homogeneous primordial seed
field inside a galaxy remains for the time being out of the scope of
this paper.

%\subsection{Modeling the magnetic field evolution}

\subsection{Initial magnetic spots: configuration and evolution
of the radial and azimuthal ordering scales}
\label{ssec:spots}

We start the description of the model from phase 3 when the disk is
formed ($t=0$). Assuming that the magnetic field in the disk is
produced by the turbulent dynamo, we start with an rms turbulent
magnetic field $b= 6.2\, \mu$G assuming equipartition with kinetic
energy of turbulence, and whose scale radius is $l_0=100$\,pc
\citep{arshakian09}. Then the average amplitude of the regular field
over the scale $l_0$ at $t=0$ is given by
\begin{equation}
B_0 \simeq b N_1^{-1/2} \simeq 0.02 \mu\mathrm{G},
\label{eq:b0}
\end{equation}
where $N_1 = 3hR^2/(2l_0^3) \simeq 75,000$ is the number of
turbulent cells in the volume of the disk for a galaxy with Gaussian
radius scale of $R=10$\,kpc and Gaussian height scale $h\approx
500$\,pc. We assume that the regular field in initial magnetic spots
is oriented randomly in the plane of the disk - the vectors
$\vec{B}_0$ change randomly from one spot of scale radius $l_0$ to
another.

The regular magnetic field is spatially ordered on a coherence scale
of $l_0=100$\,pc. The radial size of a magnetic spot grows with time
due to dynamo action. An approximate estimate of the ordering scale
of the spot in radial direction is given by
\begin{equation}
l_{r}(t) = l_0 + V_{r}t,
%\sqrt{\frac{\Omega l_0^2 v}{3h}}=l_0\left(1+t\sqrt{\frac{\Omega v}{3h}}\right),
\label{eq:lr}
\end{equation}
where $t$ is the age of a galaxy from the disk formation epoch, and
$V_{r}$ is the speed of the magnetic field propagation of the spot in radial direction. 
According to \cite{moss98}, it
depends on the radial turbulent diffusivity ($\beta_{r}=l_{0}v/3$)
and the dynamo growth rate ($\Gamma=\Omega l_0 /h$):
\begin{equation}
V_{r} = \sqrt{\beta_{r}\Gamma} = l_{0}\sqrt{\frac{\Omega v}{3h}},
\end{equation}
where $\Omega$ and $h$ are the angular rotation and scale height of
the disk, and $v$ is the turbulence velocity of the gas \citep[see]
[and references therein]{arshakian09}. \\

The turbulent cells as well as the magnetic spots are stretched
in azimuthal direction by differential rotation \citep{ruzmaikin88},
i.e. $r$ and $r+l_{r}$ move with different rotation rates
$\Omega(r)$ and $\Omega(r+l_r)$. However, the effect of turbulent
diffusion in course of dynamo action works against stretching of a
spot in azimuthal direction in the way that the pitch angle ($p$) of
magnetic lines remains constant. We suppose that the turbulent
diffusivity prevents further stretching of the spot if the ratio
$l_{\phi}/l_r$ becomes larger then $\tan^{-1} p$, i.e. if the spot
diagonal becomes more inclined than the magnetic line. Assuming that
$\Omega(r) - \Omega(r+l_{r}) \approx \Omega(r) l_{r}/r$ we arrive to
the following ordering scale in azimuthal direction

\begin{equation}
\displaystyle{
l_{\phi}(t)=
\left\{
\begin{array}{ll}
\displaystyle
l_{r}(t)\left(1+\frac{\Omega(r) t}{2}\right)  &   \\
\\ \displaystyle
l_{r}(t)\tan^{-1} p, \,\mbox{ if } \left(1+\frac{\Omega(r) t}{2}\right) > \tan^{-1} p. & 
\end{array}
\right.
}
\label{eq:la}
\end{equation}

We stress that the relations for radial and azimuthal ordering
scales are derived from the assumptions of our qualitative model and
that the ordering scales are likely to be model dependent. A
specification and improvement of scaling laws on the basis of
numerical simulation of a particular model of the galactic dynamo is
highly desirable.

The timescales of ordering the field on scales of 1\,kpc in vertical
and radial directions are comparable ($\approx 0.8$\,Gyr; Eqs.~(10)
and (11) in Arshakian et al. 2009). We start the modeling of
magnetic spots at the age of 0.8\,Gyr from the scale radius of
$L=0.5$\,kpc (scale length of 1\,kpc) which has a size of the full
height of the galaxy disk, $1$\,kpc. At this epoch, only a fraction
of spots will contribute to the dynamo-driven large-scale magnetic
field. Thus the number of useful spots of a scale radius
$L=0.5$\,kpc are those which have a magnetic pitch angle of the
growing large-scale magnetic field, $p_0\pm \delta p$, where $\delta
p$ is some tolerance, and the directions of small-scale magnetic
fields in cells $\vec{b}$ are uniformly distributed in pitch angle
$p$. The fraction of useful spots is $2\delta p/2\pi$, so that the
number of useful spots is
\begin{equation}
n \simeq 2N_2\frac{\delta p}{\pi},
\label{eq:nofs}
\end{equation}
where the factor 2 allows for the fact that both $\vec{B}$ and
$-\vec{B}$ can seed the large-scale dynamo if $\vec{B}$ has a pitch
angle of $p\pm \delta p$, and $N_2\simeq(R/L)^2=400$ is the number
of all spots in the disk. Experience with fitting azimuthal modes to
the observed magnetic fields indicates that the magnetic pitch angle
has, typically, a tolerance range $\delta p \simeq 5^{\circ}$. This
estimate can be used, rather arbitrarily, in our case. Taking these
numbers, we get $n\simeq 20$. Thus, about 20 magnetic spots have the
pitch angles within $15^{\circ}$ and $25^{\circ}$ and constitute a
regular field of $B(t)$ averaged over a radial scale of $1$\,kpc.
According to the evolutionary model \citep{arshakian09} the regular
field of $B(t)=0.6\,\mu$G will be achieved in $t\approx 0.8$\,Gyr
(or at $z\sim4.7$) if the thin disk formed at $z=10$ (see
Sect.~\ref{sec:link} and Table~\ref{tab:simipirm}).

\subsection{Evolutionary equations for turbulent and regular fields}
\label{sec:evolution}

During the evolution of an undisturbed (by major mergers) disk
galaxy with a slowly changing SFR, the turbulent field ($b$) remains
almost unchanged, while the regular field is amplified to the
equilibrium level $B(t_{\rm eql})$ (where $t_{\rm eql}$ is the time
when the strength of the regular field reaches the equilibrium
level) in a few Gyr and remains at this level \citep{arshakian09}.
Given the total magnetic field strength at the age $t$, the
amplitudes of the turbulent and regular fields can be measured from
the set of equations
\begin{equation}
B_{\rm tot}(t)^2=b^2+B(t)^2,
\label{eq:Btot}
\end{equation}
\begin{equation}
b/B(t) = \sqrt{N_1}\exp{(-t/t^{*})},
\label{eq:boB}
\end{equation}
where $t^{*}=h/(\Omega l)$ is the dynamo timescale for amplification
of the regular field \citep[][ Eq.~(9)]{arshakian09}, and
$\sqrt{N_1}=b/B(t=0)$ (see Eq.~(\ref{eq:b0})) is the number of
turbulent cells in the disk of a galaxy at $t=0$ and depends on the
ratio of geometrical volumes of the disk and the turbulent cell. At
the epoch of a disk formation, the ratio $b/B(t=0)\approx300$ for a
disk galaxy with $R=10$~kpc, $h=500$~pc and $l=100$~pc, and
$b/B(t\approx13.5 \mbox{ Gyr})=1.7$ for a contemporary Milky-Way
type galaxy (MW) at the present epoch. If the age of a galaxy is
smaller than the time when the regular field reaches the equilibrium
level, $t_{\rm eql}=t^{*}ln(2/\sqrt{N_1})$ \citep[see Eq.~(17)
in][]{arshakian09}, then we derive
\begin{equation}
b=B_{\rm tot}(t)\left[ 1+\left( \frac{e^{{t}/{t^{*}}}}{\sqrt{N_1}}
\right)^2 \right]^{-\frac{1}{2}},
\label{eq:b}
\end{equation}
and
\begin{equation}
B(t)=B_{\rm tot}(t) \frac{e^{t/t^{*}}}{\sqrt{N_1}} \left[ 1+\left( \frac{e^{t/t^{*}}}{\sqrt{N_1}} \right)^2
\right]^{-\frac{1}{2}},
\label{eq:Bt}
\end{equation}
If $t\ge t_{\rm eql}$, we assume that $b/B(t)\approx 2$ and hence
\begin{equation}
b=\frac{2} {\sqrt{5}} B_{\rm tot}(t),
\label{eq:bteql}
\end{equation}
\begin{equation}
B(t)=\frac{B_{\rm tot}(t)} {\sqrt{5}}.
\label{eq:Btteql}
\end{equation}

\subsection{Topology of the magnetic field} 
The spatial structure of the total galactic magnetic field at any age is modeled
as a superposition of a regular field ${\bf \emph{B(t)}}$ in the disk and
a random field ${\bf \emph{b}}$ which describes the contribution of
the large-scale galactic turbulence.

To describe the regular magnetic field $\vec B$ we apply the
following parametrization,
\begin{eqnarray}
B_r(r,\phi,w)    &=&{\bar B}(r,\phi,t) \exp \left[-\left(\frac{w}{h_{\rm mf}}\right)^2 \right] \sin{p}, \\
B_\phi(r,\phi,w)&=&{\bar B}(r,\phi,t) \exp \left[-\left(\frac{w}{h_{\rm mf}}\right)^2 \right] \cos{p}.
\end{eqnarray}
The amplitude of the magnetic field in the middle plane
\begin{eqnarray}
{\bar B}(r,\phi,t) = B(t)\exp{\left\{-\left[\frac{(\phi-\phi_0)r_0}{l_{\phi}(t)}\right]^2 \right\}} \times
  \nonumber \\
     \exp{\left[-\left(\frac{r-r_0}{l_{r}(t)}\right)^2\right]}
     \exp{\left[-\left(\frac{r}{r_{\rm mf}}\right)^2\right]} \tanh \frac{r}{2},
\label{midl}
\end{eqnarray}
where $r_0, \phi_0$ are coordinates of the position of the
magnetic-field spot in the disk, $B(t)$ is the amplitude of 
the regular field at the age $t$ (Eqs.~(\ref{eq:Bt},\ref{eq:Btteql})), 
$l_{r}(t)$ and $l_{\phi}(t)$ are the
radial and azimuthal ordering scales of the regular field at $t$ 
(Eqs.~(\ref{eq:lr}) and (\ref{eq:la})), $r_{\rm
mf}$, $h_{\rm mf}$ are the Gaussian radius and vertical scales of
the magnetic galactic disk, and $p$ is the pitch angle of the
magnetic field lines. We exploit the Cartesian coordinates
($x,y,w$) or cylindrical coordinates ($r, \phi, w$; $x= r \cos
\phi$, $y =r \sin \phi$). The coordinates of the galaxy projected on
the sky are ($x, y' =y \cos i$), where $i$ is the inclination angle
of the galaxy.

The first two exponential terms in Eq.~(\ref{midl}) describe the radial and
azimuthal profiles of the spot, respectively, the forth term means that
there is a cut-off in spot distribution nearby the outer border of
the galaxy, and the last one stands to make the field amplitude
vanishing just in the galactic center to make it smooth at this
exceptional point.

We found the vertical magnetic field component $B_w$ in the
disk from the solenoidility condition
\begin{equation}
\frac{\partial B_w}{\partial w}=-\frac{1}{r}
\left(\frac{\partial r B_r}{\partial r}+\frac{\partial B_\phi}{\partial \phi}\right).
\label{sol}
\end{equation}

The turbulent magnetic field is considered as a divergence free,
random fluctuating field with the given energy spectra $\Psi$ and
shape function $\Pi$:

\begin{equation}
\vec {b}(r,\phi,w)=b\,\Psi(r,\phi,w)\,\Pi(r,\phi,w)
\end{equation}
where $b$ is the strength of the turbulent field (Eqs.~(\ref{eq:b},\ref{eq:bteql})) and 
\begin{equation}
\textstyle{
\Pi(r,\phi,w)=\exp{\left[-\left(\frac{r}{r_{\rm mf}}\right)^2\right]}\exp{\left[-\left(\frac{w}{h_{\rm mf}}\right)^2\right]},
}
\end{equation}
%\begin{eqnarray}
%\vec {B}^{turb}(r,\phi,z)=\Psi(r,\phi,z)\Pi(r,\phi,z)  \\
%\Pi(r,\phi,z)=\exp{\{-(\frac{r}{\sqrt{2}R^t_g})^2\}}\exp{\{-(\frac{z}{\sqrt{2}h^t_0})^2\}}
%\end{eqnarray}
where the spectral property of the random function $\Psi$
are specified as follows:
\begin{eqnarray}
|\hat{\Psi}(\vec k)|^2 =
\left\{
\begin{array}{ll}
\left( k/k_0 \right)^{\alpha}, & k>k_0 \\
\left( k/k_0 \right)^{\beta}, & k<k_0.
\end{array}
\right.
\end{eqnarray}
For the sake of definiteness, we adopt $\alpha = -5/3$ Kolmogorov
scaling), $\beta = 2$, $k_0 = 5 h_{\rm mf}^{-1}$ \citep{stepanov08}.

\subsection{Other assumptions of the model}
\label{ssec:assumptions}

The assumptions of the model for the magnetic field evolution is
described in \cite{arshakian09}. An average gas density
of a protogalactic halo of $\approx 6\times
10^{-24}$ g\,cm$^{-3}$ within its virial radius of about 600\,pc  
was assumed at $z=15$. In the case of 
equipartition between magnetic
and kinetic energy of turbulence of 10\,km\,s$^{-1}$, the gas
densities of nearby disk galaxies varies between $2\times 10^{-24}$~g\,cm$^{-3}$ and
$18\times 10^{-24}$~g\,cm$^{-3}$, corresponding to the range of
$\sim(5-15)~\mu$G for the strength of the random magnetic field
\citep{beck05b}. For simplicity, we assume a gas density of
$\rho = 10^{-23}$ g\,cm$^{-3}$ that is unchanged during the evolution.

For a singular isothermal sphere, the rotation curve of the halo is
given by $V_{\rm c}=10 H(z) r_{200}$, where $V_{\rm c}$ is the
circular velocity of the dark halo at a limiting radius of $r_{200}$
within which the mean mass density is 200 times the background
density at the redshift $z$, and $H(z)$ is the Hubble constant at
redshift $z$ \citep{mo98}. If the gravitational effects of the disk
itself are neglected (the thin-disk rotation follows the rotation of
the halo) then the angular rotation curve of the disk is
\begin{equation}
\Omega = \frac{V_{\rm c}}{r}=\frac{10 H(z) r_{200}}{r}.
\end{equation}
In the disk-halo model \citep{mo98} $r_{200}=r_{c}\sqrt{2}/0.04$ and
$V_{\rm c}=350H(z)r_{\rm c}$ where $r_{c}$ is the scale radius
of the total cold-gas component (molecular and neutral hydrogen;
H$_2$+HI) which has an exponential distribution in the disk.
As a first approximation, we assume that $H(z)=H(0)=70$\,km s$^{-1}$
Mpc$^{-1}$ is constant, and that the angular rotation of the
Milky-Way type galaxy has a unique curve at all epochs,
\begin{equation}
    \Omega [ \mbox{km } \mbox{sec}^{-1}  \mbox{kpc}^{-1}]= \frac{24.5 r_{\rm c}}{r}.
\end{equation}

According to observations, the exponential scale radius of the
thermal emission (proportional to $n_e^2$) is about the same as the
exponential scale radius of the synchrotron intensity
\citep[e.g.][]{walsh02}, so that the exponential scale radius of the
thermal electron density $n_e$ is related to that of the synchrotron
intensity $l_{\rm syn}$ as $l_{n_e} = 2 l_{\rm syn}$. We assume that
the exponential scale radius of the total cold gas in a disk galaxy
is $l_{\rm c}$(H$_2$+HI$) =5$\,kpc, which is similar to $l_{\rm
syn}$ and the scale radius of the thermal gas $l_{\rm th}$, $l_{\rm
c}=l_{\rm syn}=l_{\rm th}=5$\,kpc. The scale radius of the electron
density $n_e$ is $l_{\rm n_e}=10$\,kpc ($l_{\rm n_e^2}=l_{\rm th}$).
In this paper, we use the Gaussian scale radius which is
$r=l\sqrt{\ln 2}$ for distributions with the same half-power width.
The relations between the Gaussian scale radii of the synchrotron
intensity $r_{\rm syn}$, cosmic ray $r_{\rm cr}$, and the magnetic
field $r_{\rm mf}$ (which is assumed to be the scale radius of all
field components: total, regular and turbulent) follow from energy
equipartition between magnetic fields and cosmic rays ($B_{\rm
tot}^2\propto n_{\rm cr}$)
\begin{eqnarray}
& &r_{\rm c} = r_{\rm syn}=r_{n_e^2}=4.2\,{\mbox{kpc}}, \nonumber \\
& &r_{n_e} = \sqrt{2}  \,r_{n_e^2}=5.9\,{\mbox{kpc}}, \nonumber \\
& &r_{\rm cr} =  \sqrt{2}  \,r_{\rm syn} = 5.9 \,{\mbox{kpc}}, \nonumber \\
& &r_{\rm mf} = 2  \,r_{\rm syn} = 8.3 \,{\mbox{kpc}}.
%r_{\rm Btot} = \sqrt{2} \, r_{\rm cr} =\sqrt{3+\alpha} \,r_{\rm syn} \simeq 2 \, r_{\rm syn}.
\end{eqnarray}

The exponential scale height of the Reynolds layer of ionized gas
($n_e$) is about 1\,kpc in the Milky Way \citep{savage09}. As this
is hard to measure in external galaxies, we assume this height to be
universal. This value is about one tenth of the exponential scale
radius, and we may assume the same ratio for the Gaussian scales
\citep{stepanov08}. The exponential scale height of synchrotron
emission is about 1.8\,kpc for the thick disk and 300\,pc for the
thin disk \citep{krause09}. Dynamo action occurs in both components.
The average value of about 0.5\,kpc is again about one tenth of the
exponential scale radius of synchrotron emission. Hence, Gaussian
scale heights can be assumed to be equal to one tenth of the scale
radius for all components.

We further assume for evolving MW-type disk galaxies ($\ga
10^{10}$\,M$_{\sun}$) that (a) the turbulence scale and turbulence
velocity were driven by SN explosions and had values close to those
observed in present-day disk galaxies such as the MW, $l=100$\,pc
and $v=10$\,km\,s$^{-1}$; (b) the scale height of the disk at
present epoch is $h=0.5$\,kpc and the scale length is $R=10$\,kpc,
the disk is thin ($h/R=0.05$) and the ratio $h/R$ is constant at any
epoch; (c) the scale length is constant over the cosmological epoch
(no-downsizing); (d) the pitch angle of the regular field is
unchanged during the evolution and $p=20^{\circ}$.

\section{Link between star-formation rate and amplitude of the total magnetic field}
\label{sec:link}

In our paper, we do not introduce a star-formation rate (SFR) as an additional parameter of the model but rather link it to the amplitude of the total magnetic field. This can be used to scale the total and polarized intensities modeled in Sect.~\ref{sec:simul}. 
The SFR is an important parameter which can be
determined from ultraviolet (UV), H$\alpha$ emission line, infrared
(IR) and radio surveys. The UV and optical SFR indicators are
strongly affected by dust, and therefore the IR and radio
luminosities have been used to estimate an unbiased SFR
\citep[see][and references therein]{bell03}.

\begin{figure}[htbp]
\begin{center}
\includegraphics[width=0.4\textwidth]{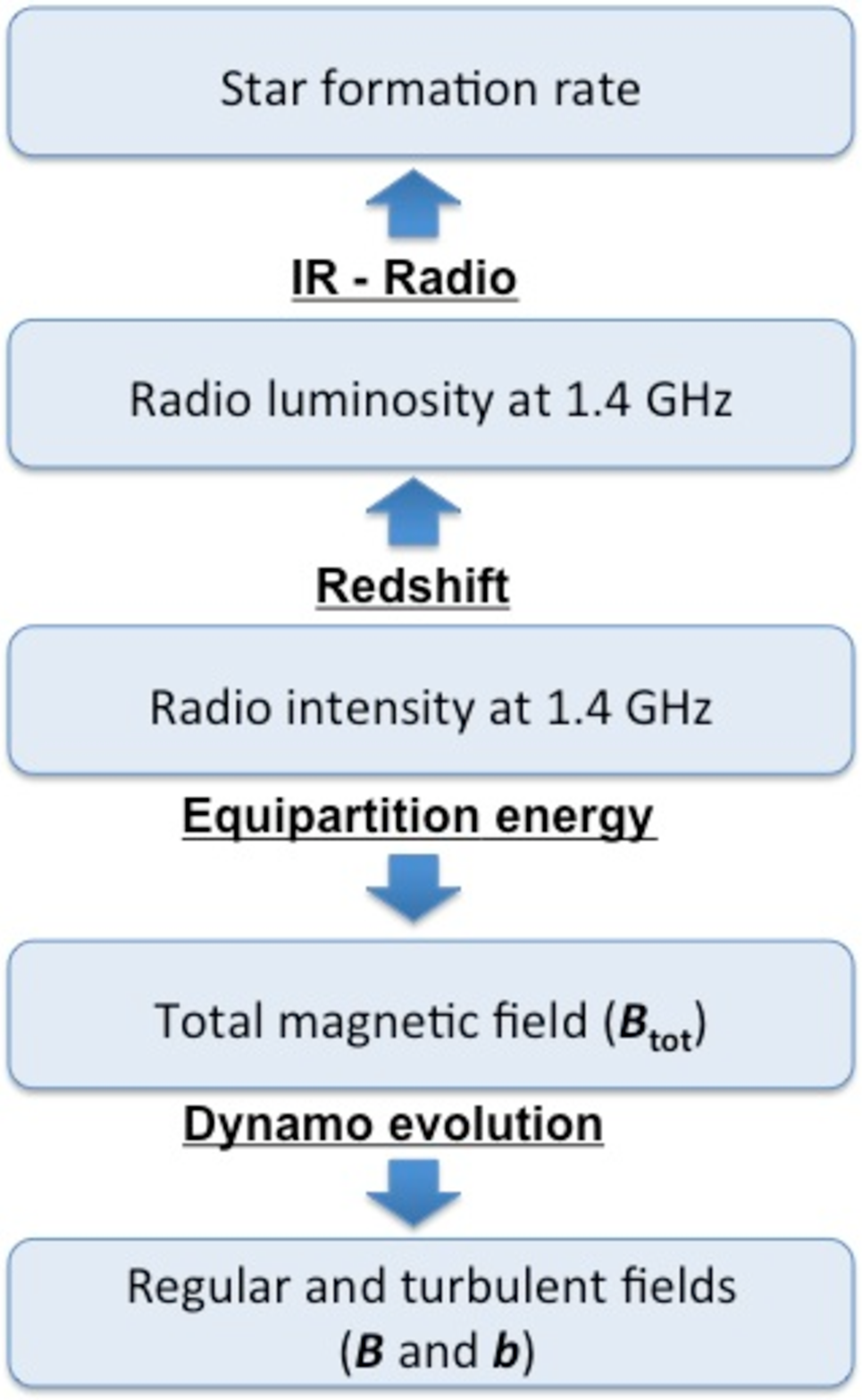}
\caption{The diagram shows the link between SFR and magnetic field
strength in a star-forming galaxy. 
%An empirical correlation between IR
%and radio emission \citep{bell03} is used to estimate the radio
%luminosity at 1.4 GHz, and then the rest frame radio intensity of a
%galaxy. The later is related to the total magnetic field strength
%\citep{beck05} under the assumption of the equipartition between the
%energies of the CRs and the total magnetic field. The regular and
%turbulent magnetic-field amplitudes are estimated from the
%total-field strength using the timescale of amplification of regular
%fields of the dynamo-evolution model \citep{arshakian09}.
}
\label{fig:diagram}
\end{center}
\end{figure}

Here, we derive the link between the SFR and the amplitude of the
total magnetic field. \cite{bell03} used radio and IR luminosities of
star-forming galaxies to calibrate the radio-derived SFRs. He showed
that IR emission is a good indicator of the star formation in
luminous galaxies, while in galaxies fainter by two orders of
magnitude the IR emission traces a small fraction of the star
formation. Because of the tight radio--IR correlation, this implies
that the radio synchrotron emission is also suppressed in
low-luminosity galaxies, compared to brighter galaxies which already
has been observed in nearby spiral galaxies \citep{dumke00,chyzy07}
and in dwarf galaxies \citep{klein91}. This is taken into account in
deriving the relation between the average SFR (in
$M_{\odot}$~yr$^{-1}$) and the rest-frame radio luminosity at
1.4\,GHz \citep[see][ Eq.~(6)]{bell03}:
\begin{eqnarray}
  \lefteqn{ \mbox{SFR(radio)}=} \nonumber \\
  %\nonumber \\
  & & =\left\{
\begin{array}{cc}
5.52\times10^{-22}L_{1.4}, &  \mbox{if } L_{1.4}>L_c \\ \\
\displaystyle
\frac{5.52\times10^{-22}L_{1.4}}{0.1+0.9(L_{1.4}/L_{c})^{0.3}}, &
\mbox{if } L_{1.4}\le L_c, \\
\end{array}
\right.
\label{eq:sfr}
\end{eqnarray}
where $L_c=6.4\times10^{21}$\,W Hz$^{-1}$.

From the observed radio intensity of a galaxy one can estimate the radio luminosity and hence the star formation rate from Eq.~(\ref{eq:sfr}). On the other hand, the radio intensity can be used to estimate the total magnetic field strength \citep{beck05} under the assumption of the equipartition between the energies of the CRs and the total magnetic field. In turn, the regular and
turbulent magnetic-field amplitudes can be estimated from the
total field strength (Eqs.~(\ref{eq:b}-\ref{eq:Btteql})) using the timescale of amplification of the regular
fields of the dynamo-evolution model \citep{arshakian09}. The scheme presented in Fig.~\ref{fig:diagram} demonstrates the link between the SFR and the amplitudes of the total magnetic field, regular and turbulent fields.

\begin{table*}[htbp]
\caption{Radio luminosities, total field strengths and
star-formation rates (SFR) in nearby spiral galaxies}
\label{tab:luminosity} \centering
\begin{tabular}{cccrrcrccc}
\hline
Source & Type & $i$ & $D$ & $S_{4.85}$ & $L_{1.4}$ & $B_{\rm tot}$ & SFR(radio) & SFR\,(IR, H$\alpha$) & Ref.\\
$ $ & $ $  & $(\degr)$ & (Mpc)  & (mJy) & (W Hz$^{-1}$) &  $\mu$G  & $M_{\odot}\,\mbox{yr}^{-1}$ & $M_{\odot}\,\mbox{yr}^{-1}$ &  \\
\hline
IC 342    & SAB(rs)cd   &   25  &   3.1  &   856   &  $2.5\times10^{21}$  &  10    &  1.6   & 1.2  &  1 \\
M 31      & SA(s)b      &   76  &   0.7  &  1863   &  $2.7\times10^{20}$  &   3    &  0.3  & 0.30 &  2 \\
M 33      & SA(s)cd     &   56  &   0.8  &  1539   &  $3.3\times10^{20}$  &   7    &  0.36  & 0.45 &  3 \\
M 51      & SAbc        &   20  &   9.7  &   380   &  $1.1\times10^{22}$  &  17    &  5.3   & 3.1  &  4 \\
M 81      & SA(s)ab     &   59  &   3.3  &   385   &  $1.2\times10^{21}$  &   8    &  0.96   & 0.89  &  5 \\
M 83      & SAB(s)c     &   24  &   3.7  &   809   &  $3.4\times10^{21}$  &  18    &  2.0   & 1.6  &  1 \\
NGC 253   & SAB(s)c     &   78  &   3.9  &  2707   &  $1.3\times10^{22}$  &  35    &  6.2   & 6.3  &  1 \\
NGC 891   & SA(s)b? sp  &   88  &   9.6  &   286   &  $8.1\times10^{21}$  &   6    &  3.9   & 3.3  &  1 \\
NGC 3628  & SAb pec sp  &   89  &   6.7  &   247   &  $3.4\times10^{21}$  &   6    &  2.0   & 1.1  &  1 \\
NGC 4565  & SA(s)b? sp  &   86  &  12.5  &    54   &  $2.6\times10^{21}$  &   7    &  1.7   & 1.3  &  1 \\
NGC 4631  & SB(s)d      &   85  &   7.5  &   476   &  $8.2\times10^{21}$  &   6    &  4.0   & 2.1  &  1 \\
NGC 5775  & SAbc        &   81  &  26.7  &    94   &  $2.1\times10^{22}$  &   8    & 10.   & 7.3  &  1 \\
NGC 5907  & Sc          &   87  &  11.0  &    72   &  $9.9\times10^{20}$  &   4    &  2.0   & 1.3  &  1 \\
NGC 6946  & SAB(rs)cd   &   38  &   7.0  &   457   &  $6.8\times10^{21}$  &  16    &  3.3   & 3.2  &  4 \\
\hline

\end{tabular}

\begin{list}{}{}
%Columns are:
\item The columns are: source name, morphological type of a galaxy,
$i$ is the inclination (degrees), $D$ is the distance of a galaxy
(Mpc), $S_{4.85}$ is the total flux density at 4.85\,GHz, $B_{\rm t}$ is
the total magnetic field strength, SFR is the star-formation rate
estimated from the radio luminosity at 4.85\,GHz, SFR\,(IR,
H$\alpha$) is the star-formation rate estimated from IR or H$\alpha$
emission, and the last column is the references to the
SFR\,(H$\alpha$, IR) values. The references to SFR are: 1: Young et
al. (1989), 2: Tabatabaei \& Berkhuijsen (2010), 3: Verley et al.
(2009), 4: Leroy et al. (2008), 5: Gordon et al. (2004).
\end{list}

\end{table*}

\begin{figure}[htbp]
\begin{center}
\includegraphics[width=0.4\textwidth,angle=-90]{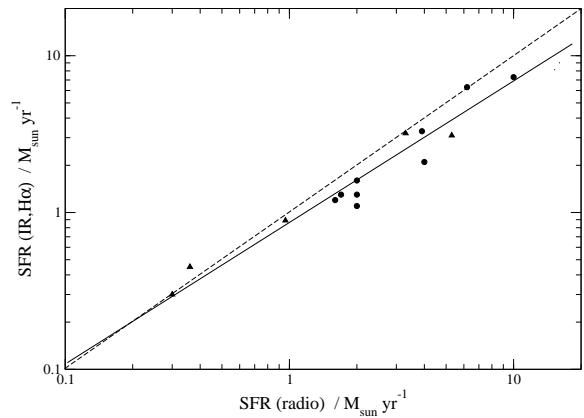}
\caption{Comparison of SFRs of 14 nearby galaxies estimated from the
radio emission at 1.4\,GHz and the IR or H$\alpha$ emission. SFRs of
nine galaxies (circles) are taken from \cite{young89} and five
galaxies (triangles) from other papers (see
Table~\ref{tab:luminosity}). The dashed line represent the equality
and the full line is the best linear fit to the data.}
\label{fig:SFR}
\end{center}
\end{figure}

To test this scheme, we selected a sample of 14 nearby spiral galaxies with different
inclinations and levels of star formation. For these galaxies, we
integrated the observed radio intensities at 4.85\,GHz and
determined their flux densities $S_{4.85}$ as described in
\cite{stil09} (see Table~\ref{tab:luminosity}). Dividing by the area
of integration gives the mean total intensities $I_{4.85}$
(mJy/beam) for these sources. We estimated the non-thermal
intensities $I_{n4.85}$ accounting for a mean thermal fraction of
20\,\%. Further, from the 4.85\,GHz flux density values $S_{4.85}$ we determined the radio luminosities $L_{4.85}$ and extrapolated these
values to 1.4\,GHz assuming a total spectral index of $\alpha=0.7$.
The values for $L_{1.4}$ are also summarized in
Table~\ref{tab:luminosity}. We used Eq.~(\ref{eq:sfr}) to calculate
the values for the SFR(radio) from the radio luminosity at 1.4\,GHz
$L_{1.4}$ (see Table~\ref{tab:luminosity}). To check the reliability of 
estimated SFR(radio), we compared them with the SFR values determined 
from the IR-emission \citep{young89} for these galaxies, adopting Eq.~(3) of
\cite{kennicutt98} or more recent values from the literature for
special galaxies. The values determined from the radio emission and
the other values correlate well for the 14 galaxies with $\rm{SFR(H
\alpha, IR)} \propto \rm{SFR(radio)}^{0.88 \pm 0.06}$ with a
correlation coefficient of 0.97 (see Fig.~\ref{fig:SFR}). Good
agreement between independent measurements suggests that the SFR can
be reliably estimated from the radio luminosity at 1.4\,GHz
(Eq.~(\ref{eq:sfr})).

Using the $I_{n4.85}$ (see Table~\ref{tab:luminosity})
and the pathlength through the emitting medium (assumed to
be the diameter of the galaxy for edge-on galaxies), we estimated
the total magnetic field strength $B_{\rm tot}$ under the assumption 
of the equipartition between the total energy of cosmic rays and that of the 
magnetic field \citep[][their Eq.~(3), using $\nu=4.85$\,GHz and a synchrotron 
spectral index of $\alpha_{n}=0.8$]{beck05}. The knowledge of the SFR of a galaxy and its effective diameter (or the pathlength through the emitting medium) allows the $B_{\rm tot}$ to be estimated, and vise versa, the SFR can be derived from the $B_{\rm tot}$ and effective radius of a galaxy. In turn, independent measurements of a SFR and $B_{\rm tot}$ open a possibility to calculate the effective diameter of unresolved sources in deep radio surveys.

In the next section, we present simulations for one specific SFR because a variation of the SFR simply scales with $B_{\rm tot}$ in the case of an unchanged size of the galaxy during the evolution, as was assumed in this work.

\section{Modeling the regular magnetic fields and radio emission properties}
\label{sec:simul}

We perform modeling for a galaxy of the radius of 10\,kpc inclined at an angle of 
$60^{\circ}$, and having a constant SFR=10\,$M_{\sun}$~yr$^{-1}$ which corresponds 
to the radio luminosity $L_{1.4} = 1.8\times10^{22}$\,W Hz$^{-1}$ at 1.4\,GHz
(Eq.~(\ref{eq:sfr})). The strength of the total magnetic field at the age $t \ge t_{\rm eql}=1.2$\,Gyr is estimated to be $B_{\rm tot} (t)= 6.9\,\mu$G, and the amplitudes of turbulent
and regular fields are $b=6.2\,\mu$G and
$B(t)=3.1\,\mu$G (Eqs.~(\ref{eq:bteql}) and (\ref{eq:Btteql})).

We start the modeling of magnetic spots at the age of 0.8\,Gyr
(Sect.~\ref{ssec:spots}). We assume that the formation of $n=15$ magnetic
spots of a scale length of 1\,kpc in the disk (with a scale radius
$8.3$\,kpc and scale height $0.5$\,kpc) are in place at $\approx
0.8$\,Gyr after the disk formation. The amplitudes of the regular
field of magnetic spots are distributed randomly around the mean
coherent magnetic field strength $B(t=0.8)=0.6\,\mu\mathrm{G}$,
within the range from $0.7 B(t=0.8)$ to $1.3 B(t=0.8)$, and coherent
fields in the spots are oriented within the pitch angles $20^{\circ}
\pm 5^{\circ}$ or $\pi + (20^{\circ} \pm 5^{\circ})$ with equal
probability.

\begin{table}[h]
\caption[]{The amplitudes of regular and turbulent
fields, radial and azimuthal coherence scales simulated for
different ages of a disk galaxy with inclination of $60^{\circ}$,
scale radius of $R=10$\,kpc, and SFR=10\,$M_{\sun}$ yr$^{-1}$. }
\label{tab:simipirm}
\begin{center}
\begin{tabular}{ccccccc}
\hline
$t^{\mathrm{a}}$  & $B$     &  $b$ & $l_r$&$l_{\phi}$ & $z^{\mathrm{b}}$ & $\nu_{\rm z}^{\mathrm{c}}$ \\
(Gyr)&$(\mu$G)&($\mu$G)&(kpc)&(kpc)&      &    (MHz) \\
\hline

0    & 0.02    &  6.2 &  0.1  &  0.1  &  10  &  1650\\
%0.5 & 0.39    &  6.2 &  0.7  &  1.9  & 5.5  & 975  \\
0.8 & 0.6      &  6.2 &  1.0  &  2.9  & 4.7  & 855 \\
1.3   & 3.1    &  6.2 & 1.6   &  4.5  & 3.6  & 690 \\
2   & 3.1       &  6.2 & 2.5   &  6.8  & 2.5  & 525  \\
3   & 3.1       &  6.2 & 3.6   & 10.   & 1.9  & 435  \\
5   & 3.1       &  6.2 & 6.0   & 16.5 & 1.1  & 315  \\
%10 & 3.1       &  6.2 & 10.6 &  XXX  & 0.4  & 205  \\
13 & 3.1       &  6.2 & 15.4 &  42.4  & 0.0 & 150  \\

\hline
\end{tabular}
\end{center}

\begin{list}{}{}
\item[$^{\mathrm{a}}$] $t$ is the age of a galaxy after the disk
formation epoch assumed to occur at $z=10$.
\item[$^{\mathrm{b}}$] $z$ is the redshift of a galaxy calculated for a flat cosmology model
($\Omega_{\Lambda}+\Omega_{m}=1$) with $\Omega_{m}=0.3$ and
$H_0=70$ km\,s$^{-1}$\,Mpc$^{-1}$.

\item[$^{\mathrm{c}}$] $\nu_{\rm z}$ is the rest frame frequency
which is detected at 150\,MHz ($\lambda=2$\,m) in the frame of the
observer.
\end{list}

\end{table}

\begin{figure*}
\includegraphics[width=0.32\textwidth]{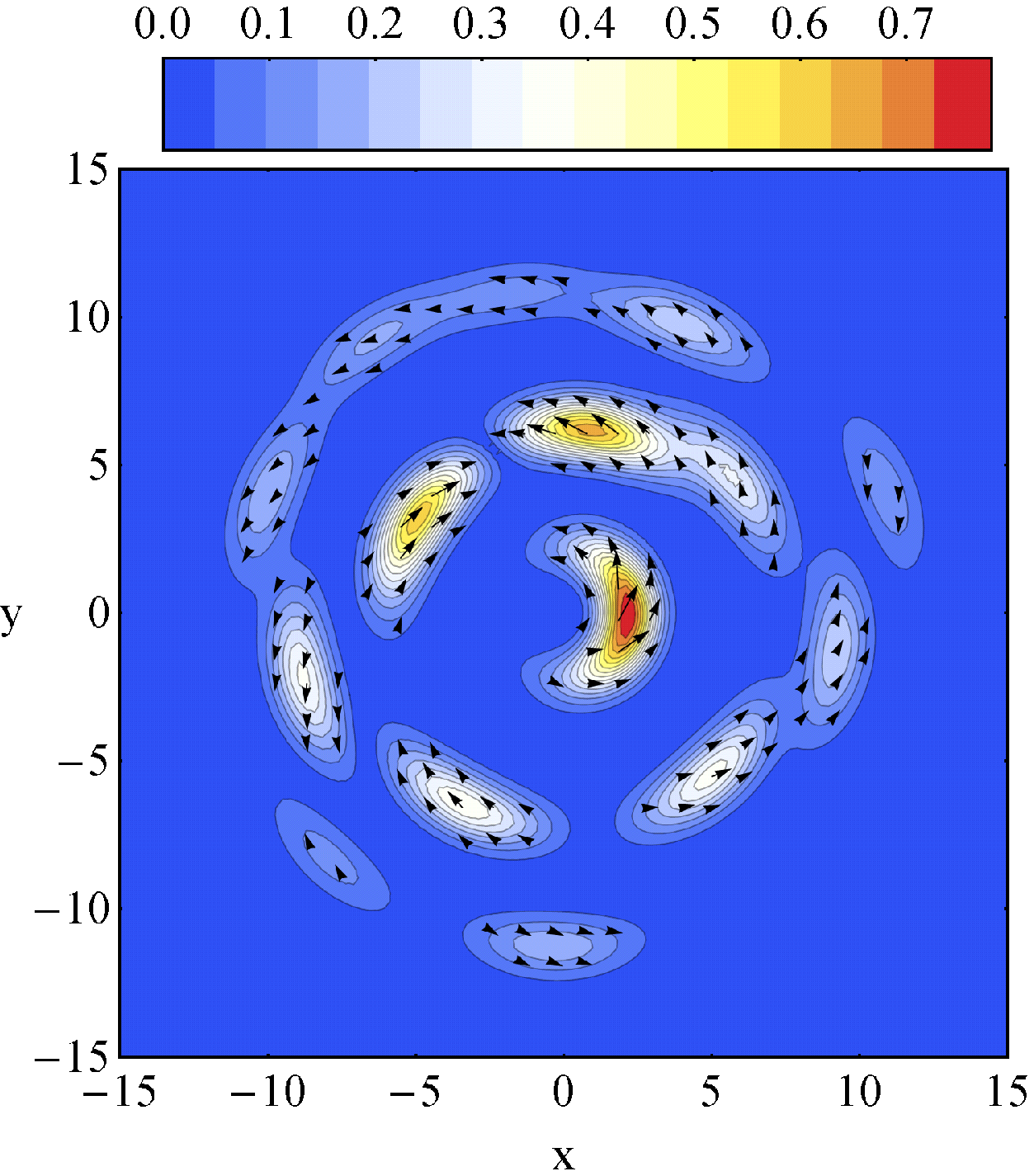}
\includegraphics[width=0.32\textwidth]{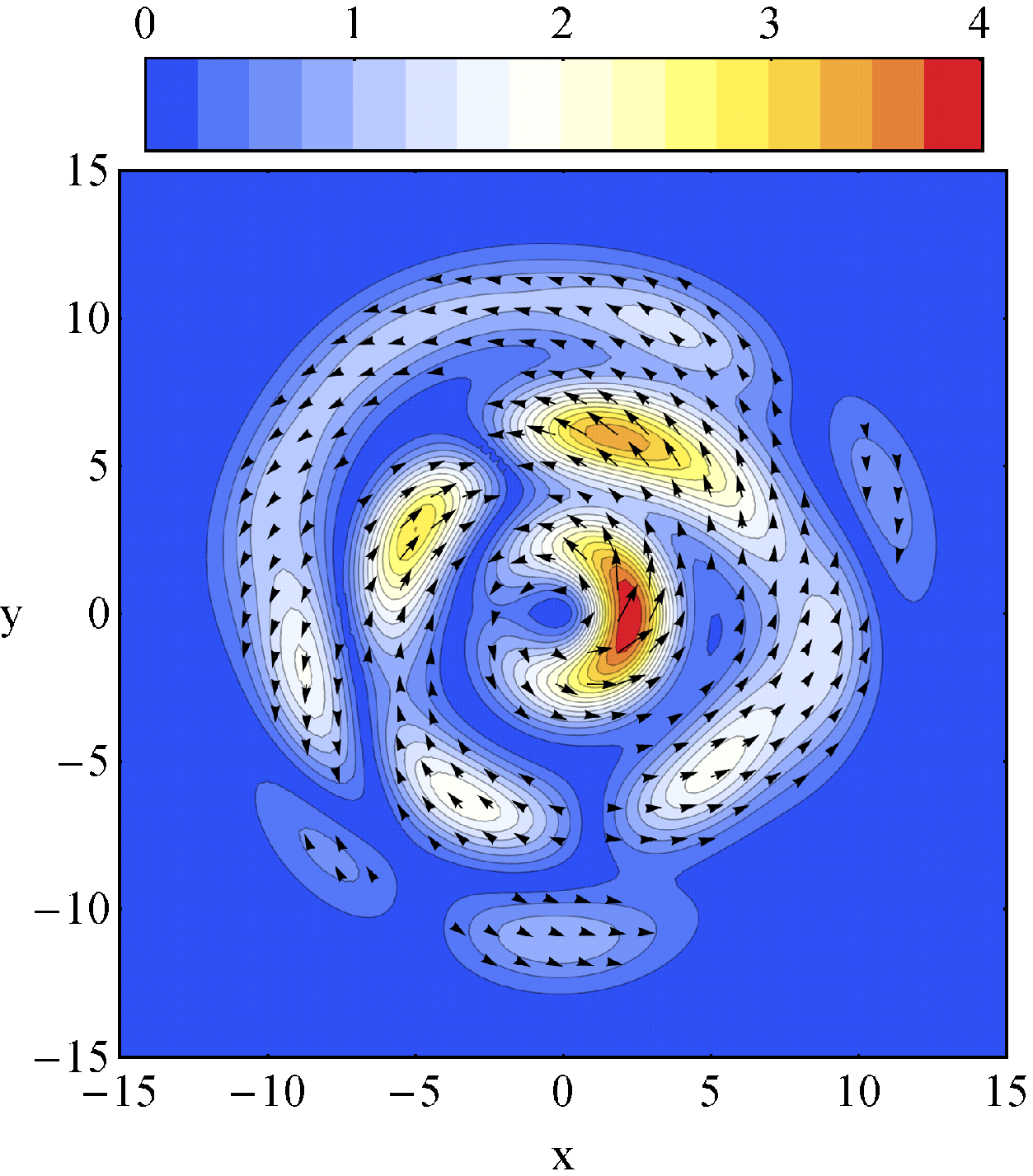}
\includegraphics[width=0.32\textwidth]{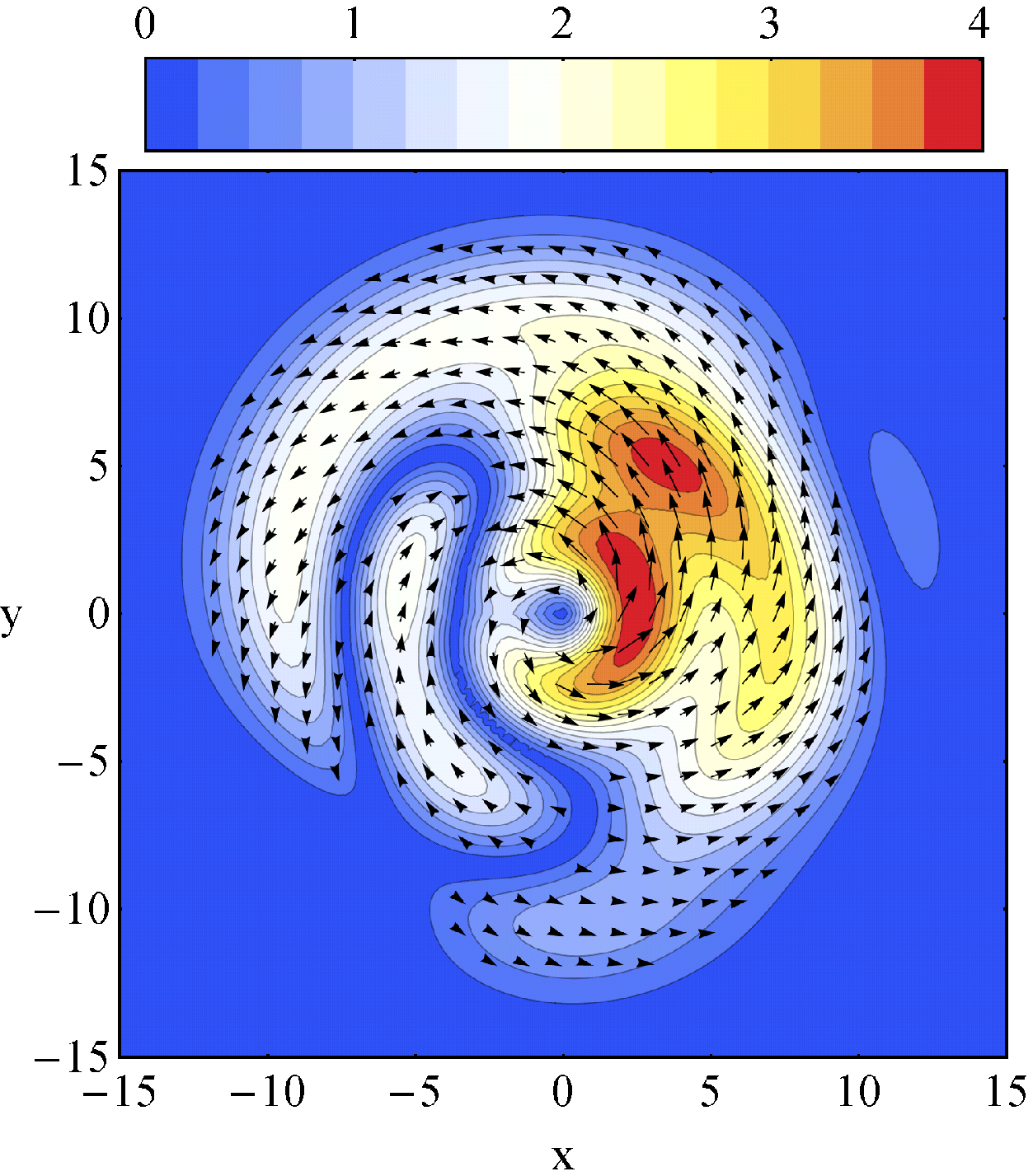}  \\
\includegraphics[width=0.32\textwidth]{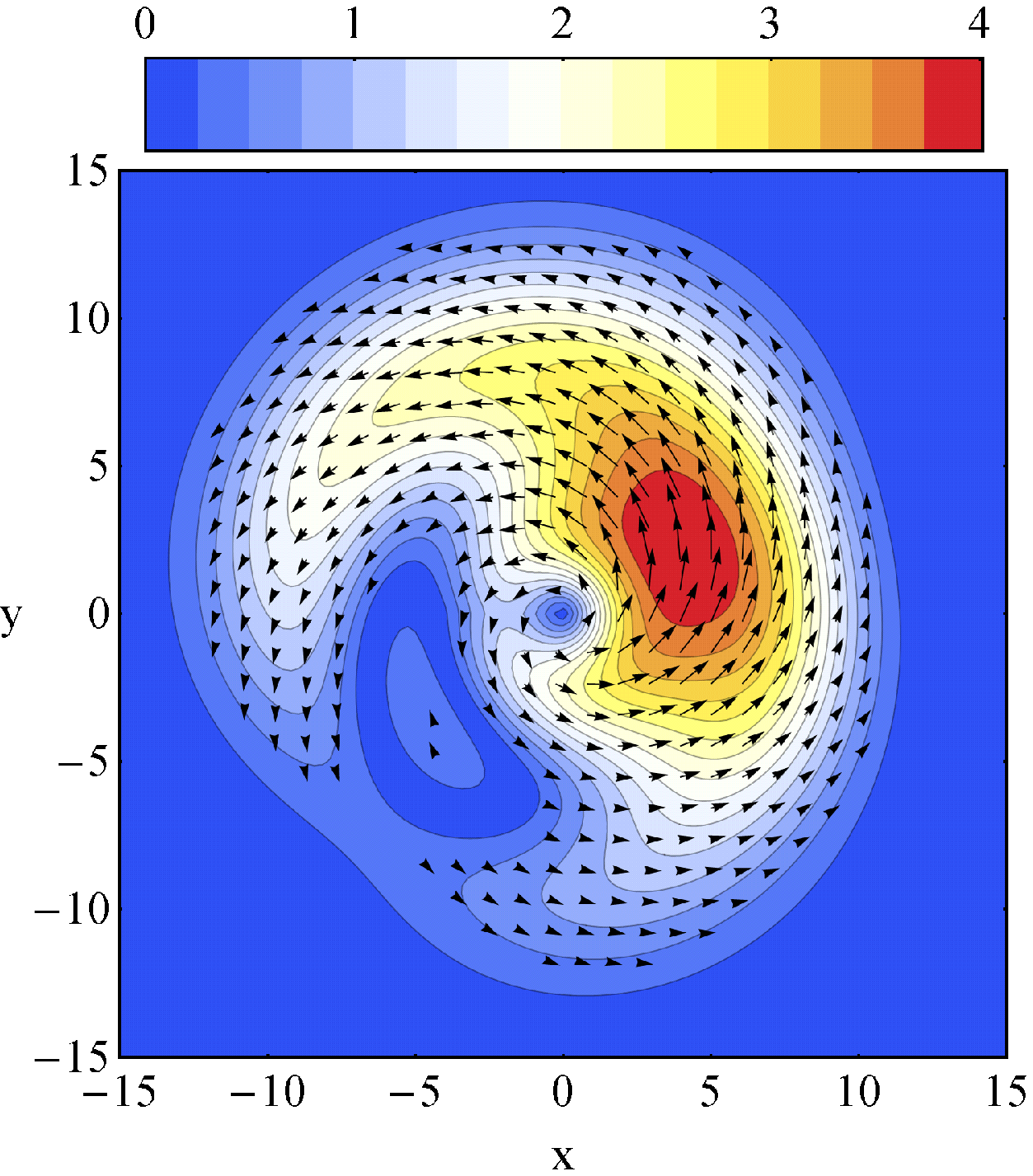}
\includegraphics[width=0.32\textwidth]{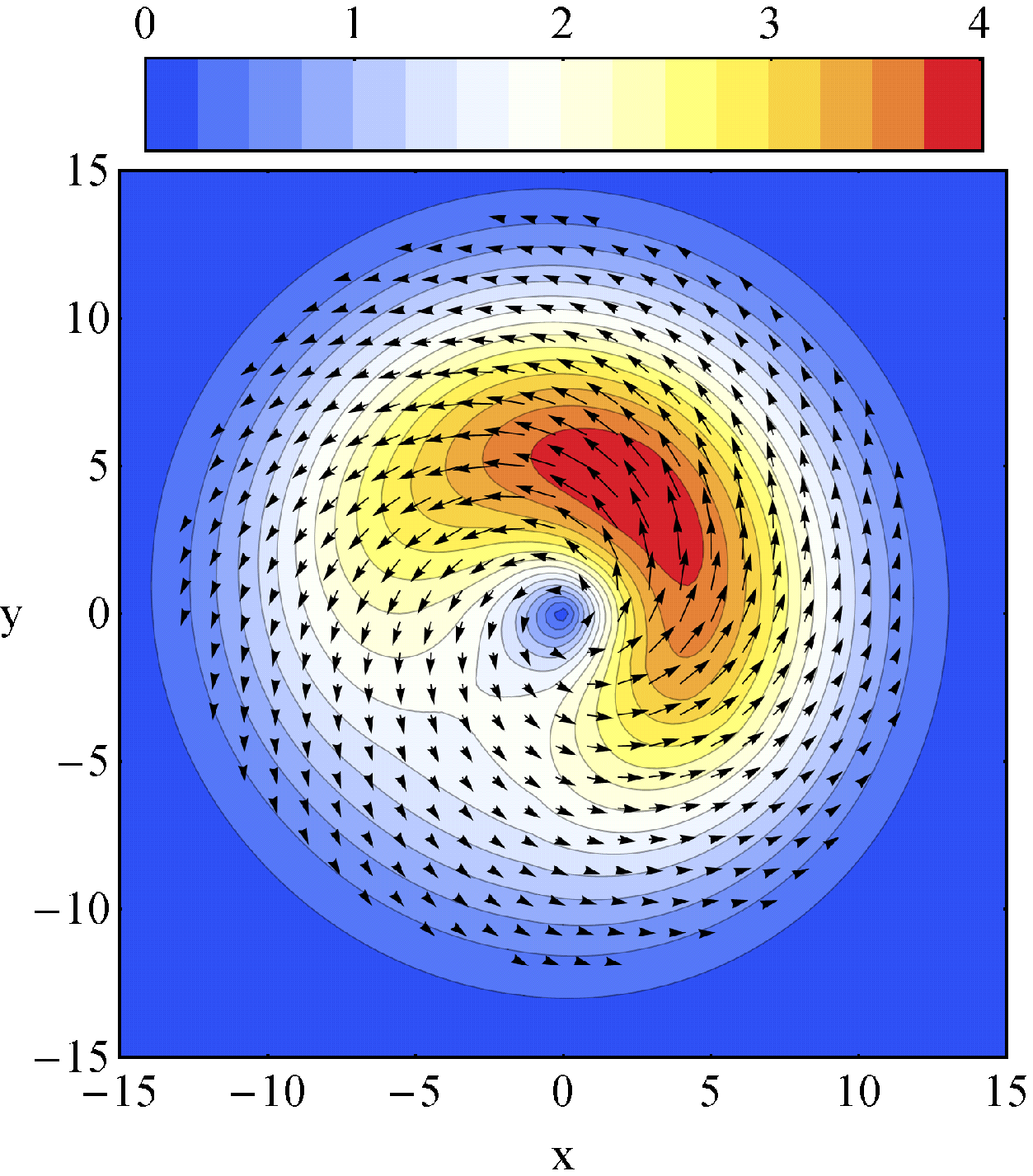}
\includegraphics[width=0.32\textwidth]{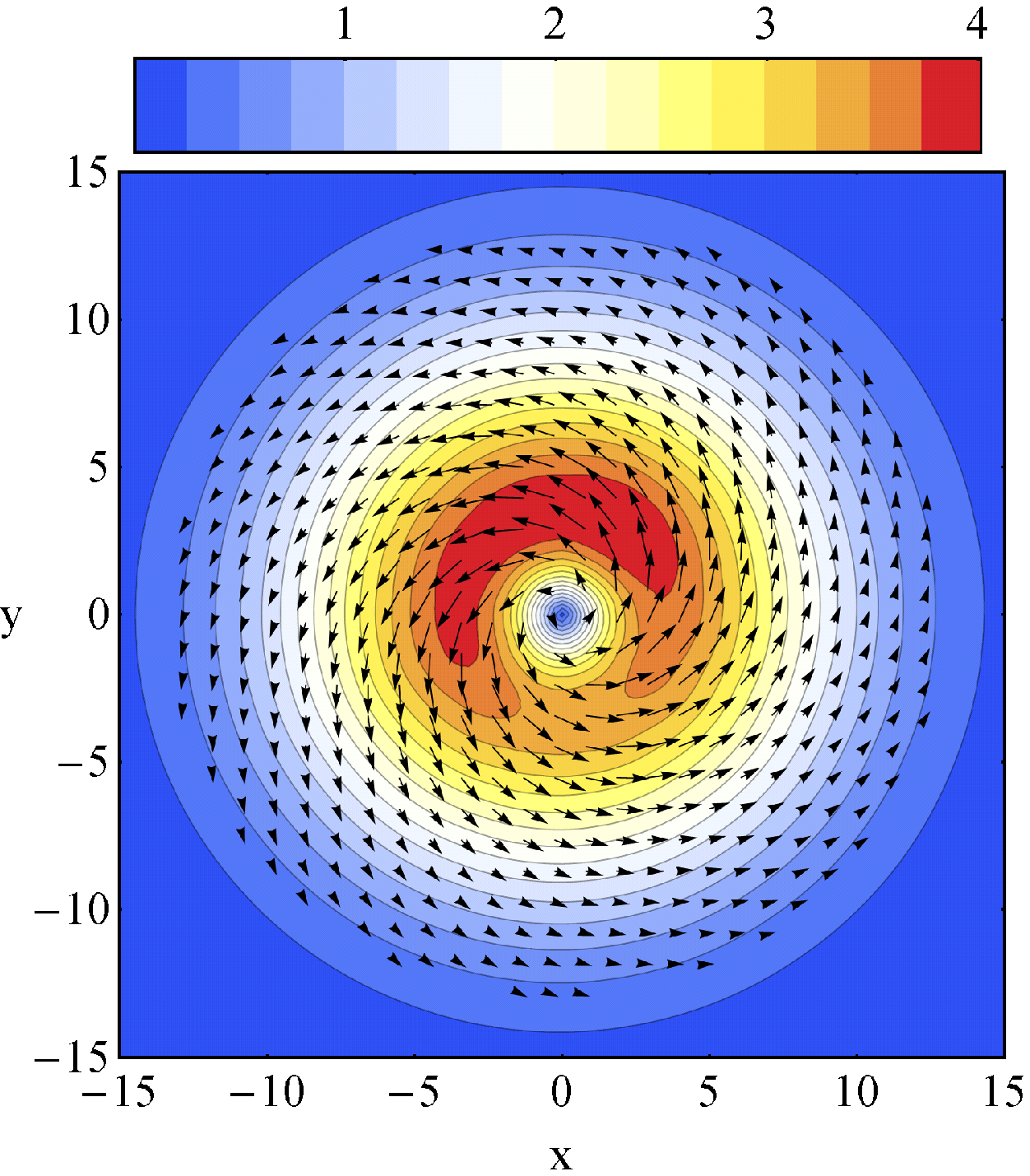}
\caption{Simualtions in the framework of the SKA Design Studies
(SKADS): the evolution of regular magnetic fields in the disk of a
galaxy seen face-on (the frame size is 20\,kpc$\times$20\,kpc). The
amplitude and ordering scale of the regular fields are estimated at
0.8\,Gyr after the disk formation epoch ($\sim 1.3$ $\mu$G and 1\,kpc), 1.3\,Gyr ($\sim 3.1$
$\mu$G and 1.6\,kpc), 2\,Gyr ($3.1$ $\mu$G and 2.5\,kpc), 3\,Gyr
($\sim 1.5$ $\mu$G and 7\,kpc), 5\,Gyr
($\sim 3.1$ $\mu$G and 6\,kpc), and after 13\,Gyr ($\sim 3.1$ $\mu$G
and 15.4\,kpc). The color bar represents the amplitude of the regular field in units of $\mu$G.}
\label{fig:simrf}
\end{figure*}

Modeling the regular fields for an undisturbed (without a major
merger) Milky-Way type galaxy (Fig.~\ref{fig:simrf}) show that the development of spotty
structures with a coherence length of $\sim 1.5$\,kpc is clearly
visible during $\la 1.3$\,Gyr after the disk formation. During the
second Gyr the spotty patterns develop to prolonged structures of
magnetic fields with a coherence length of $\sim 2.5$\,kpc with few
reversals of the field. After $\sim 5$\,Gyr the reversals disappear
and the coherence scale of the regular field reaches $\sim 6$\,kpc.
The fully coherent field is developed in $\sim 5$\,Gyr while the
asymmetry of the magnetic field strength remains longer.

\begin{figure*}
\centering

\begin{picture}(550,330)(0,0)
\put(0,240){
\includegraphics[width=0.33\textwidth]{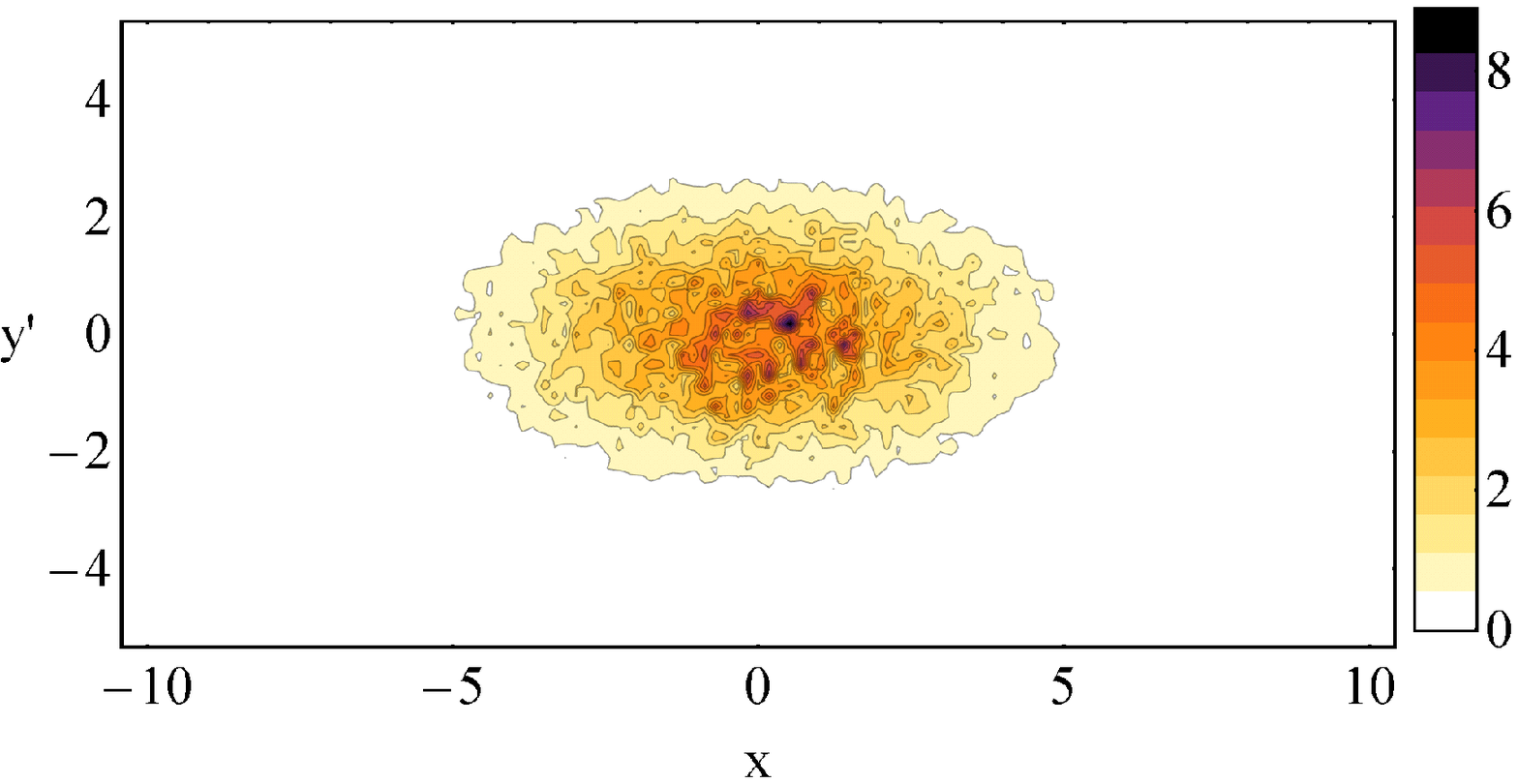}
\includegraphics[width=0.33\textwidth]{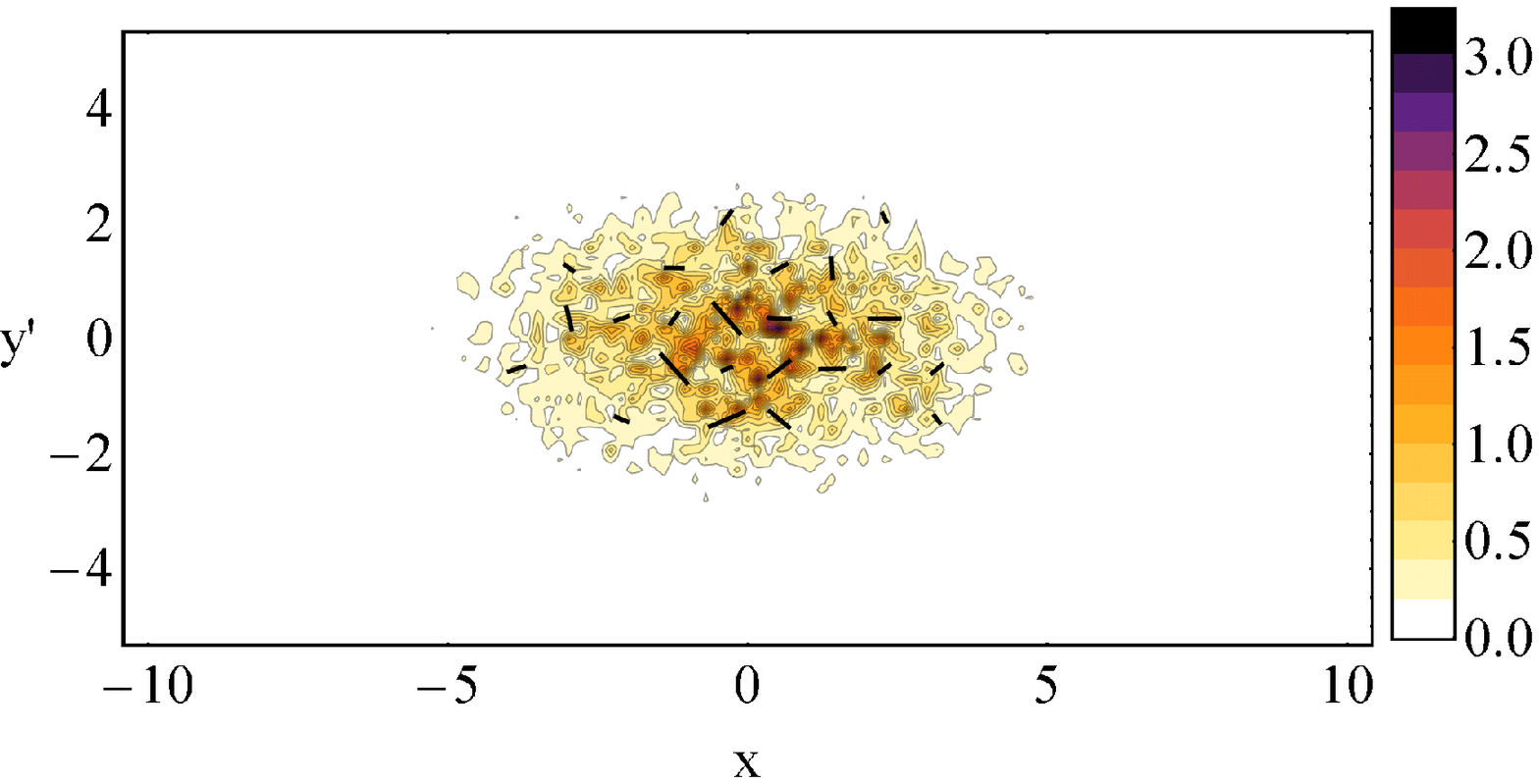}
\includegraphics[width=0.33\textwidth]{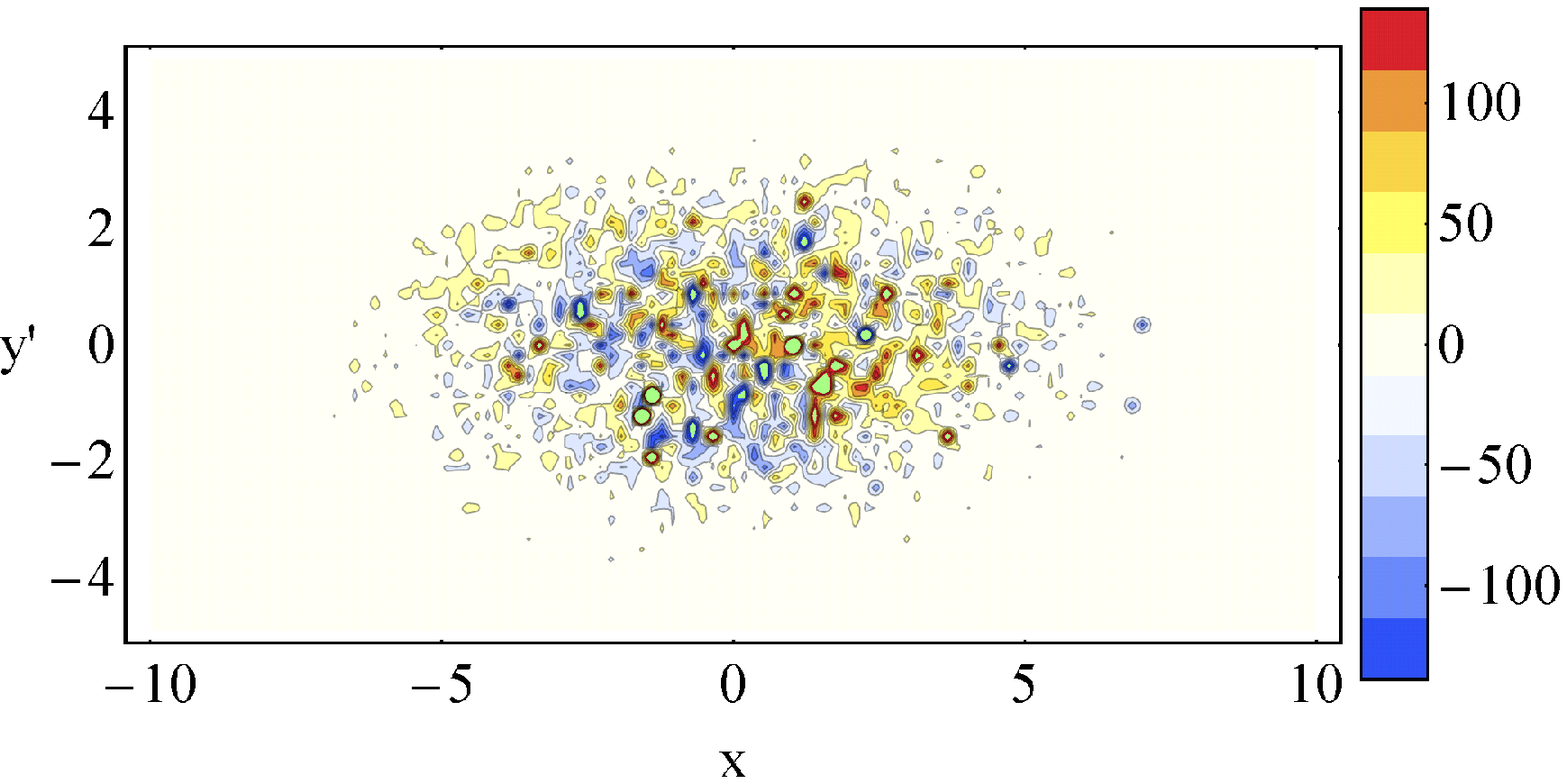}  
} 
\put(0,160){
\includegraphics[width=0.33\textwidth]{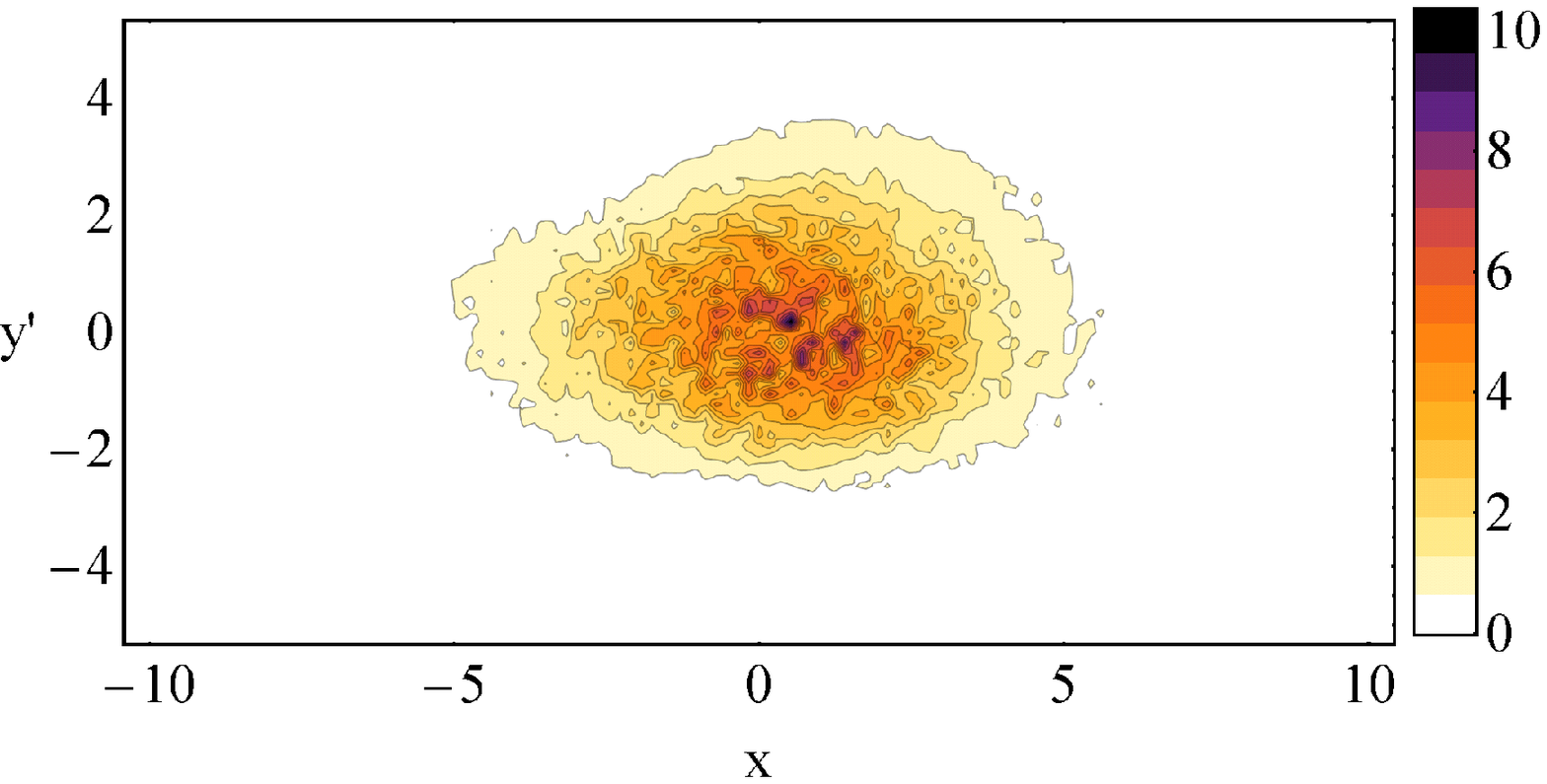}
\includegraphics[width=0.33\textwidth]{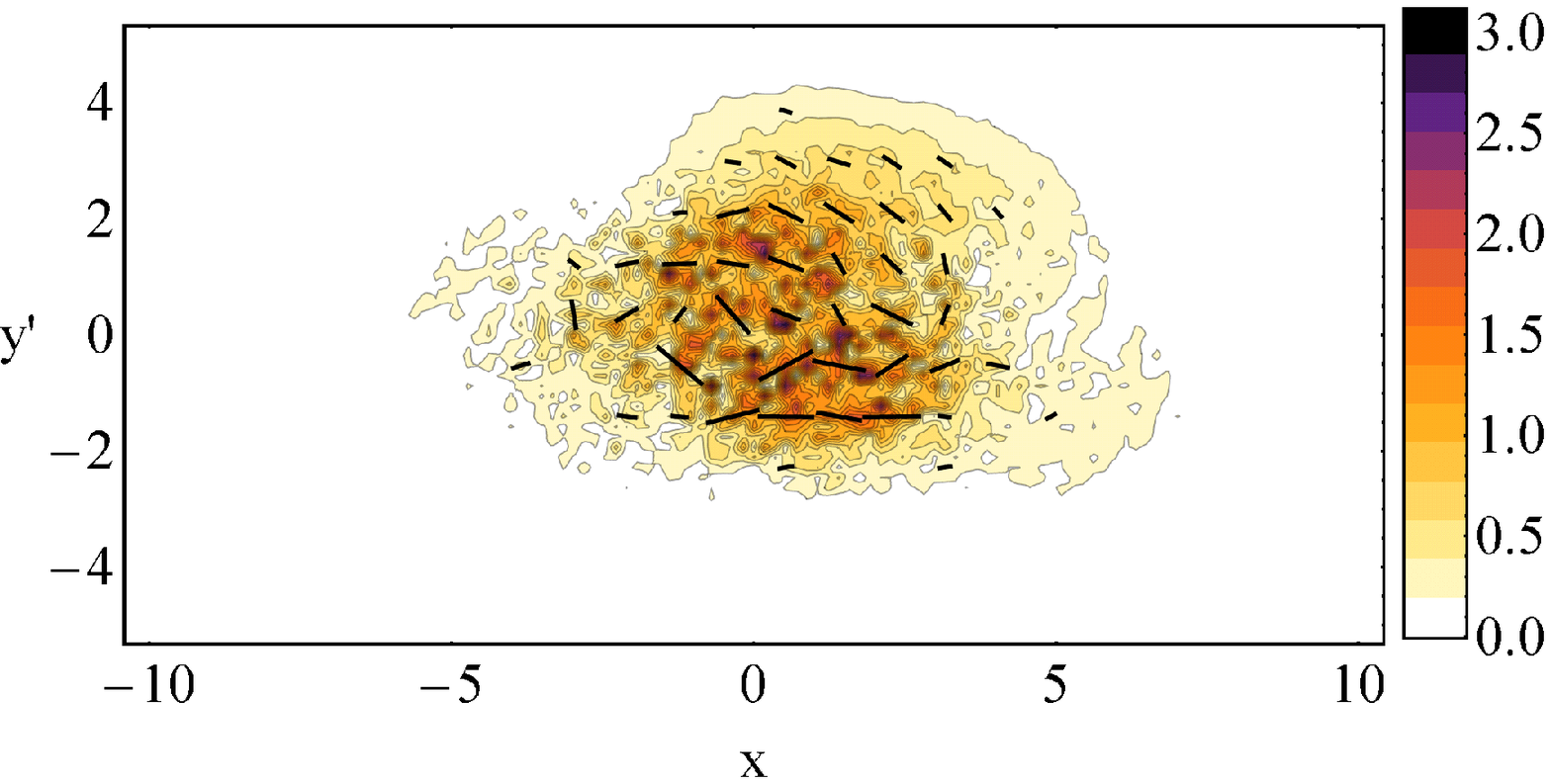}
\includegraphics[width=0.33\textwidth]{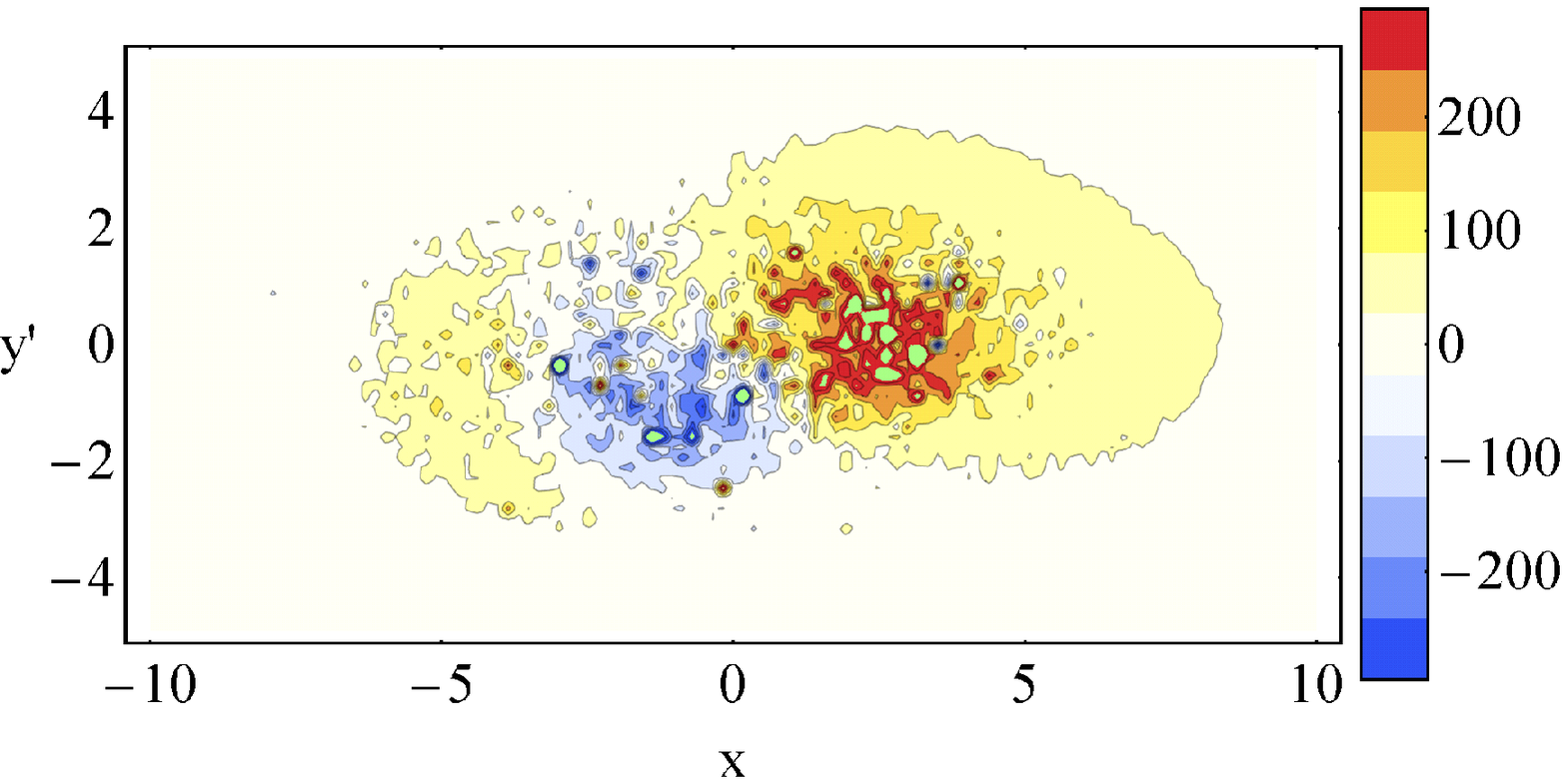}  
}
\put(0,80){
\includegraphics[width=0.33\textwidth]{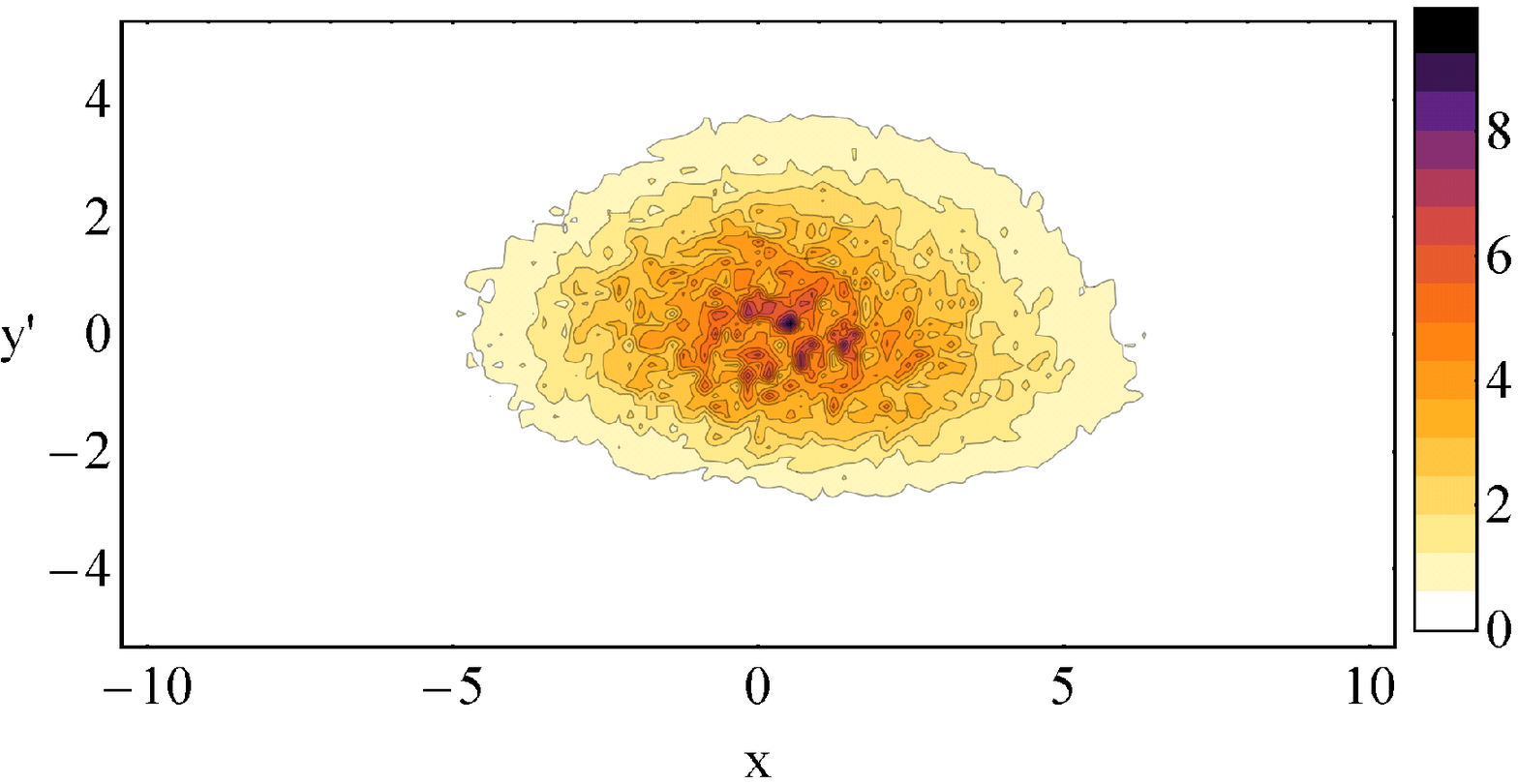}
\includegraphics[width=0.33\textwidth]{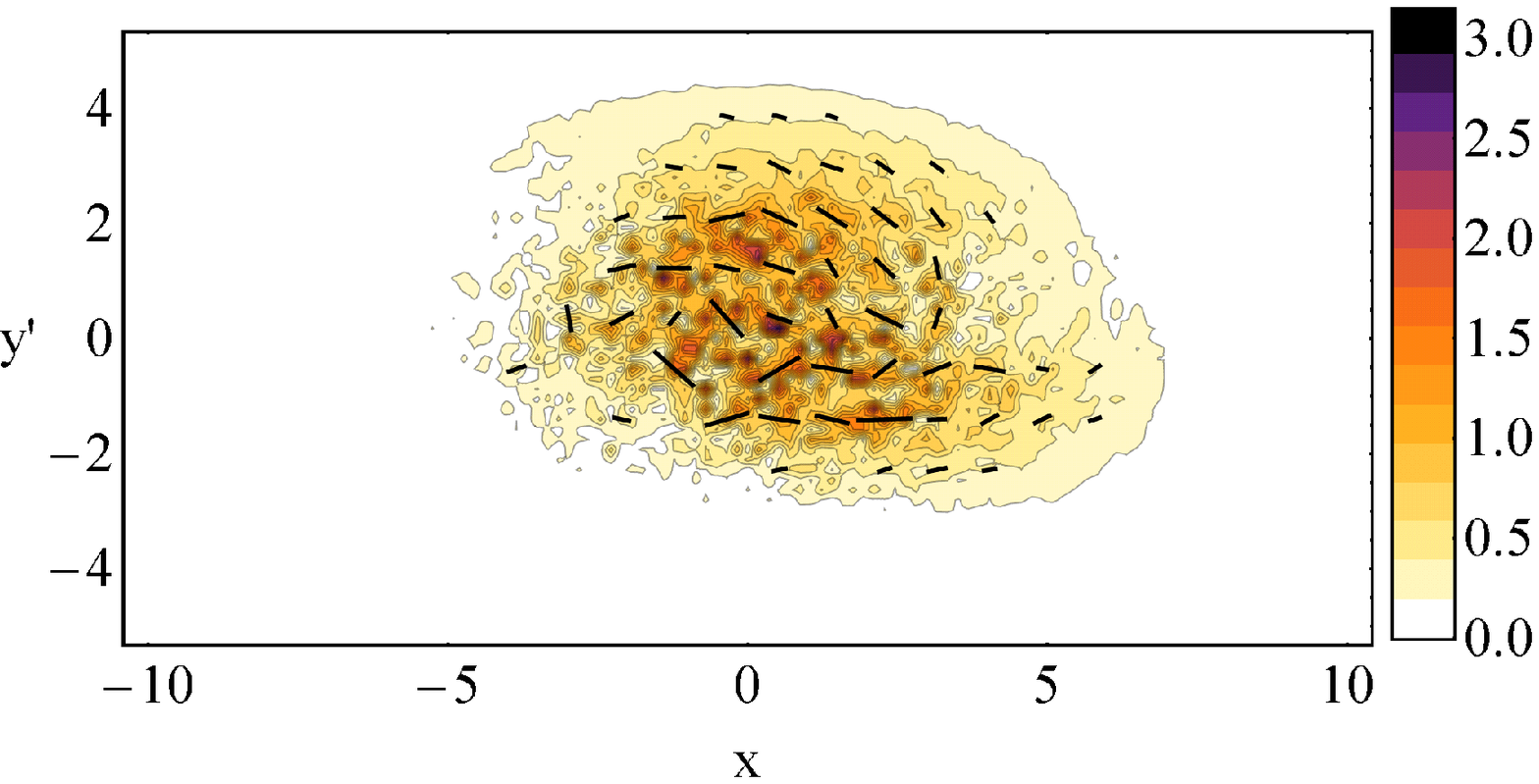}
\includegraphics[width=0.33\textwidth]{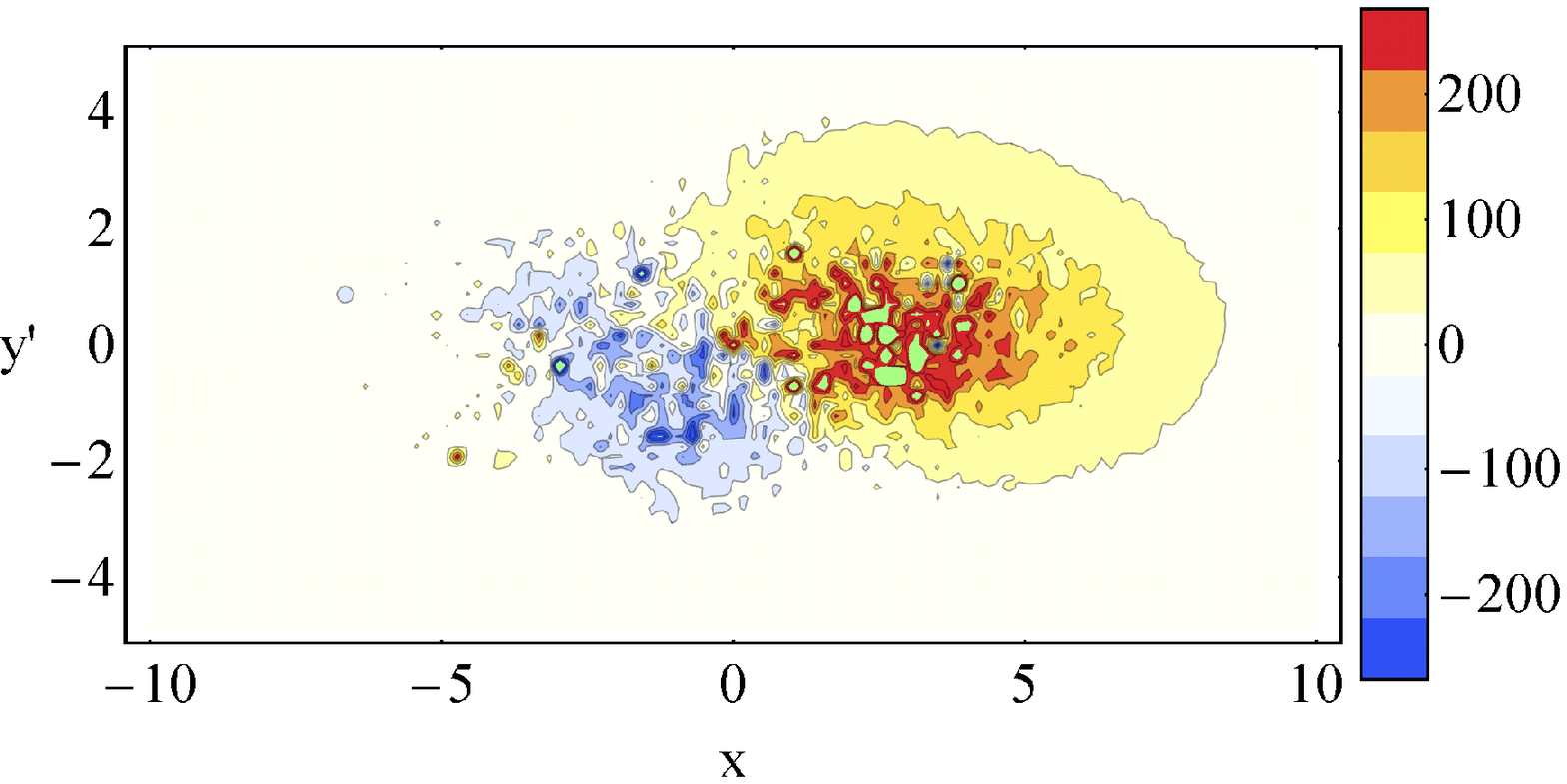}  
}
\put(0,0){
\includegraphics[width=0.33\textwidth]{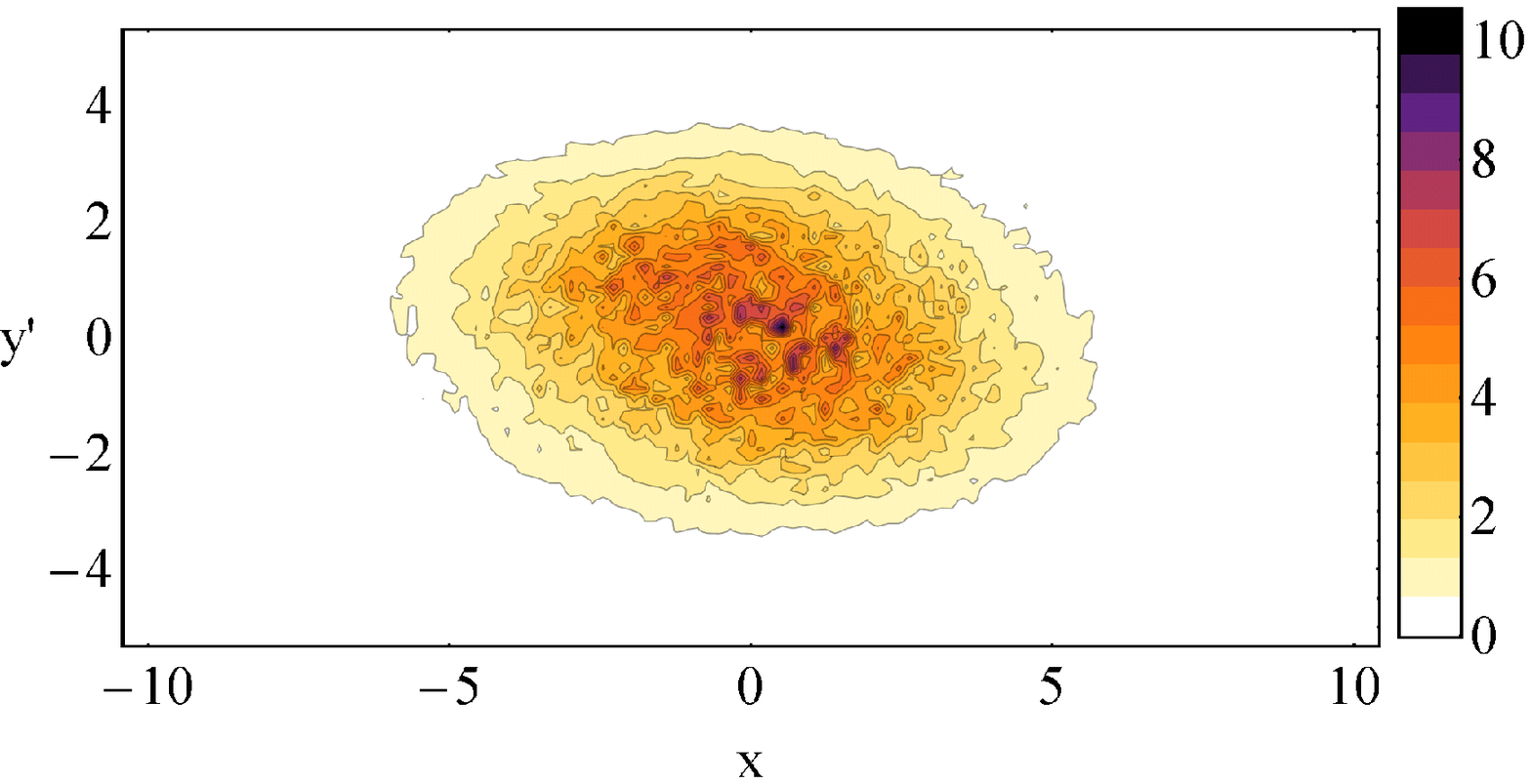}
\includegraphics[width=0.33\textwidth]{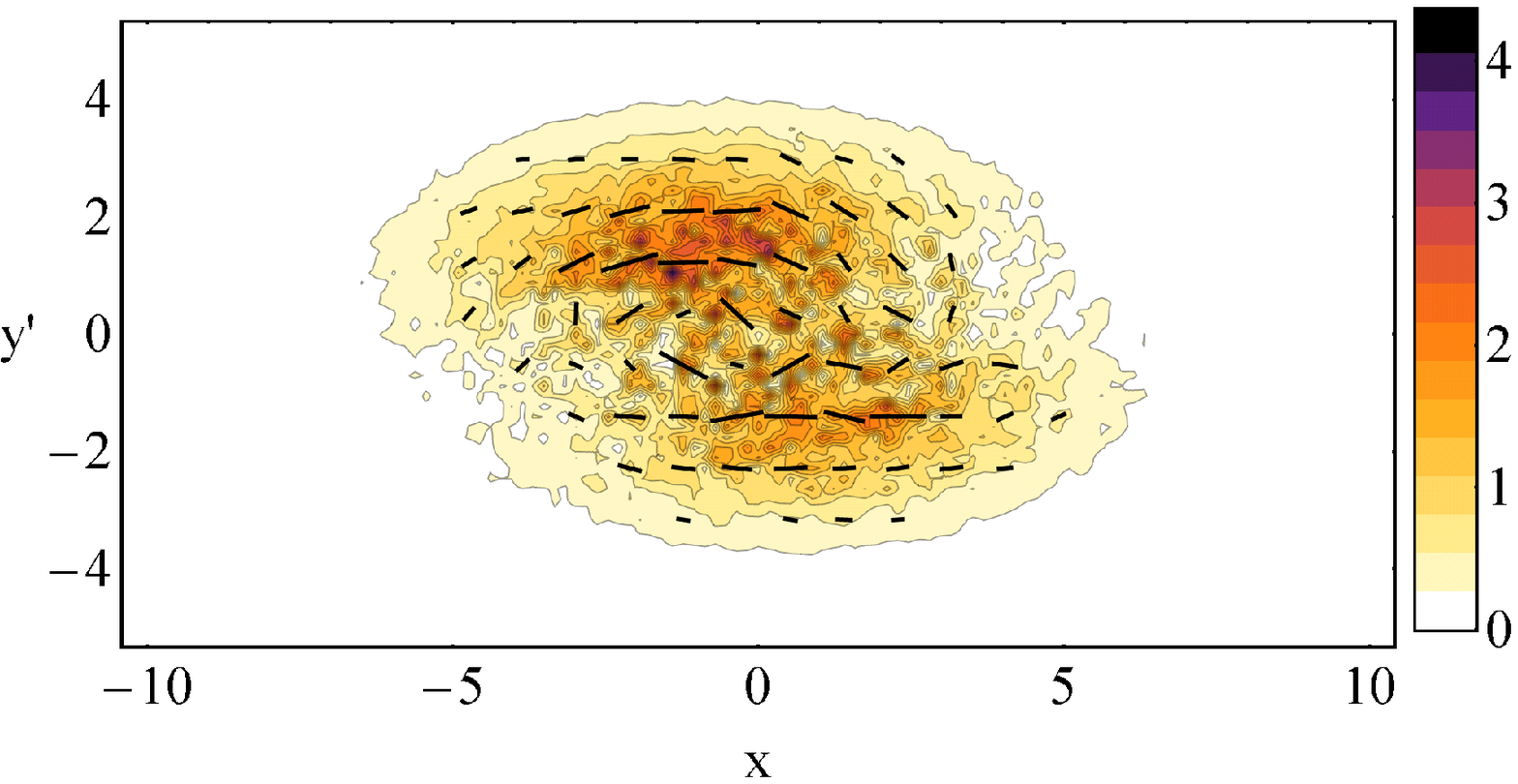}
\includegraphics[width=0.33\textwidth]{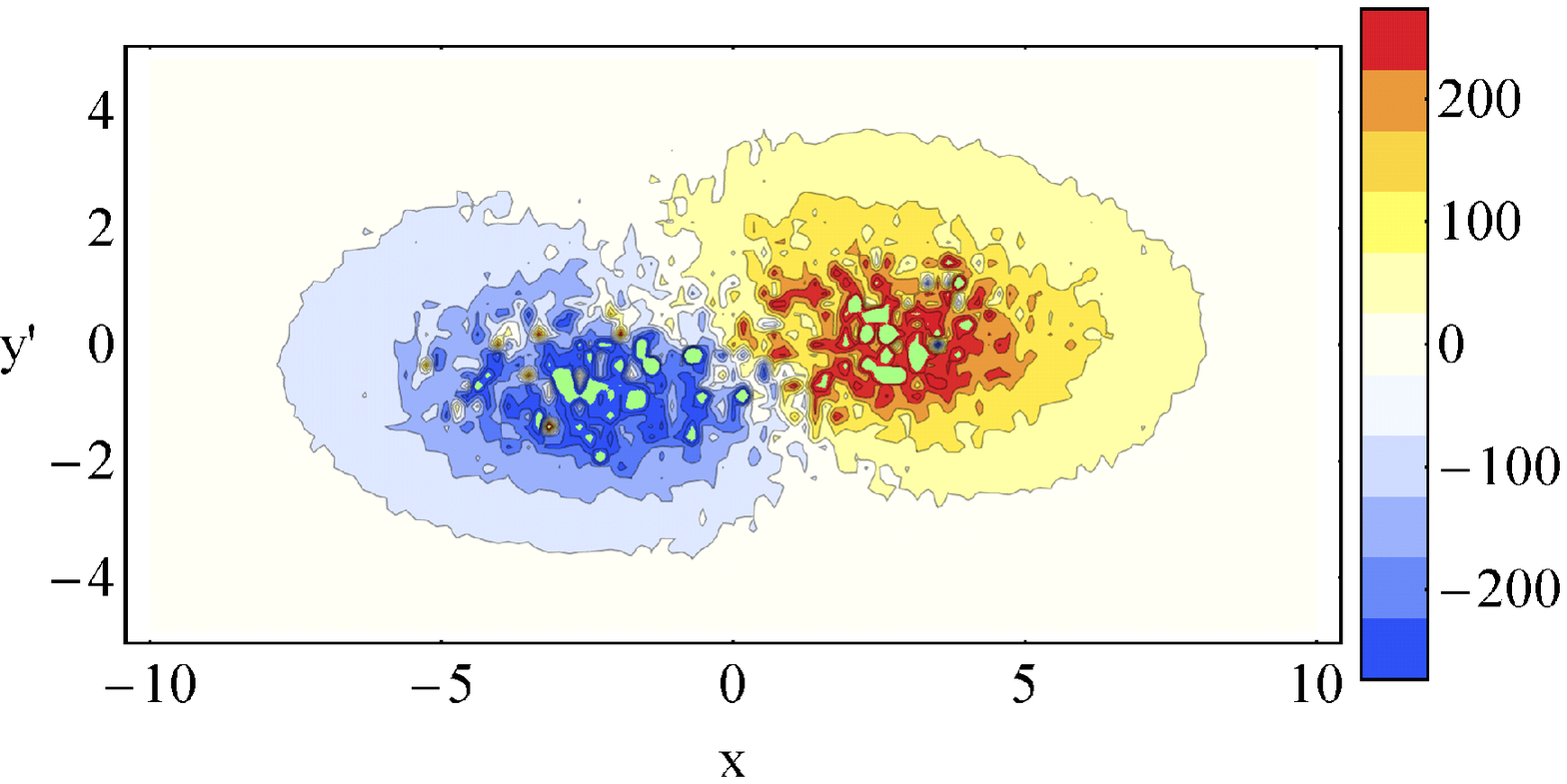} }

\put(80,325){total}
\put(220,325){polarized}
\put(400,325){RM}
\put(20,310){0.8\,Gyr}
\put(20,230){2\,Gyr}
\put(20,150){3\,Gyr}
\put(20,70){13\,Gyr}
\end{picture}

\caption{The \emph{rest frame}\ simulations of the total intensity
(\emph{left panel}), polarization (\emph{middle panel}), and Faraday
rotation (\emph{right panel}) at {\bf 5\,GHz} for a galaxy with an
inclination angle of $60\degr$, turbulent (6.2~$\mu$G) and regular
(from 1.3\,$\mu$G to 3.1\,$\mu$G, see Table~\ref{tab:simipirm})
magnetic fields, and star-formation rate of 10\,$M_{\sun}$ yr$^{-1}$
are shown for 0.8\,Gyr, 2\,Gyr, 3\,Gyr, and 13\,Gyr after disk
formation. The frame units are given in kpc. The color bars in the first and 
second columns (total and polarized intensity) are given given in arbitrary units. The color bar of the third column (Faraday rotation measure) is given in units of rad\,m$^{-2}$. } 
\label{fig:sim5ghz}
\end{figure*}

Simulations of the total intensity, polarization and Faraday depth
(see Appendix~A for a mathematical description of equations) in the
rest frame of a galaxy at 5\,GHz after 0.8\,Gyr, 2\,Gyr, 3\,Gyr, and
13\,Gyr of disk formation are shown in Fig.~\ref{fig:sim5ghz}. The
symmetric structure in total intensity at the age of 0.8\,Gyr
(Fig.~\ref{fig:sim5ghz}; first row) is due to strong turbulent field
compared to the regular field (see Table~\ref{tab:simipirm}). Smooth
and weak polarized intensity and small Faraday rotation with no
large-scale patterns are the result of a small ordering scale and
small amplitude of the regular field at this age. At the age of
2\,Gyr, the regular field has a coherence scale of $\sim 2.5$\,kpc
and few reversals, reaches the equipartition level and becomes
comparable with the strength of the turbulent field ($B\approx
b/2$). These changes are manifested by appearance of polarization
patterns and asymmetric, inhomogeneous RM structures
(Fig.~\ref{fig:sim5ghz}; second row). After 13\,Gyr, the regular
field has no reversals and is coherent at the size of a galaxy
(Fig.~\ref{fig:sim5ghz}; third row). The polarization map looks
symmetric with notable depolarization along the major axis, and the
RM structure is smooth and asymmetric. Note that the intensity units in Fig.~\ref{fig:sim5ghz} have to be scaled via SFR (or total radio luminosity; see Sect.~\ref{sec:link}).

\begin{figure*}
\centering

\begin{picture}(550,330)(0,0)
\put(80,240){
\includegraphics[width=0.33\textwidth]
{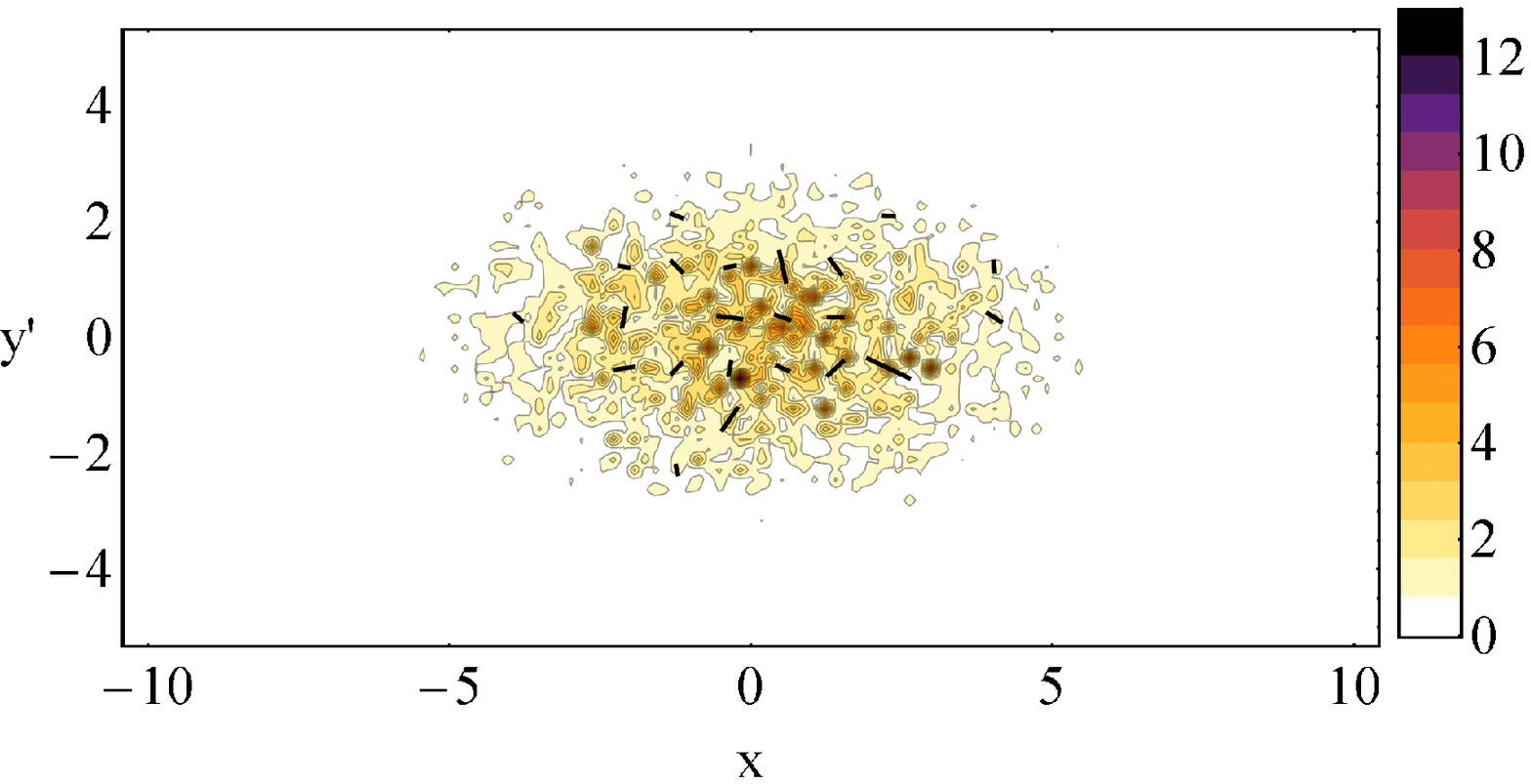}
\includegraphics[width=0.33\textwidth]
{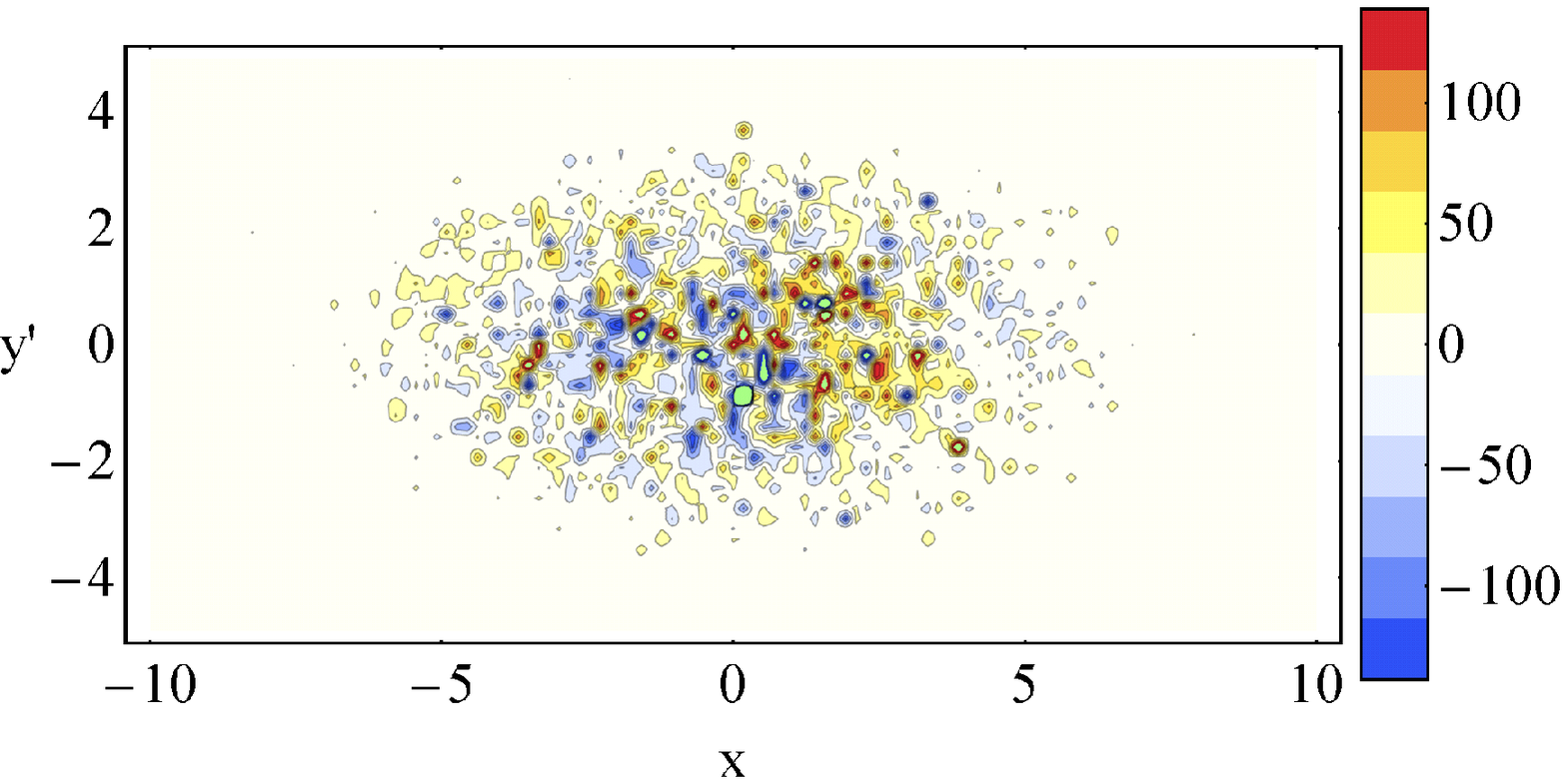}
}
\put(80,160){
\includegraphics[width=0.33\textwidth]{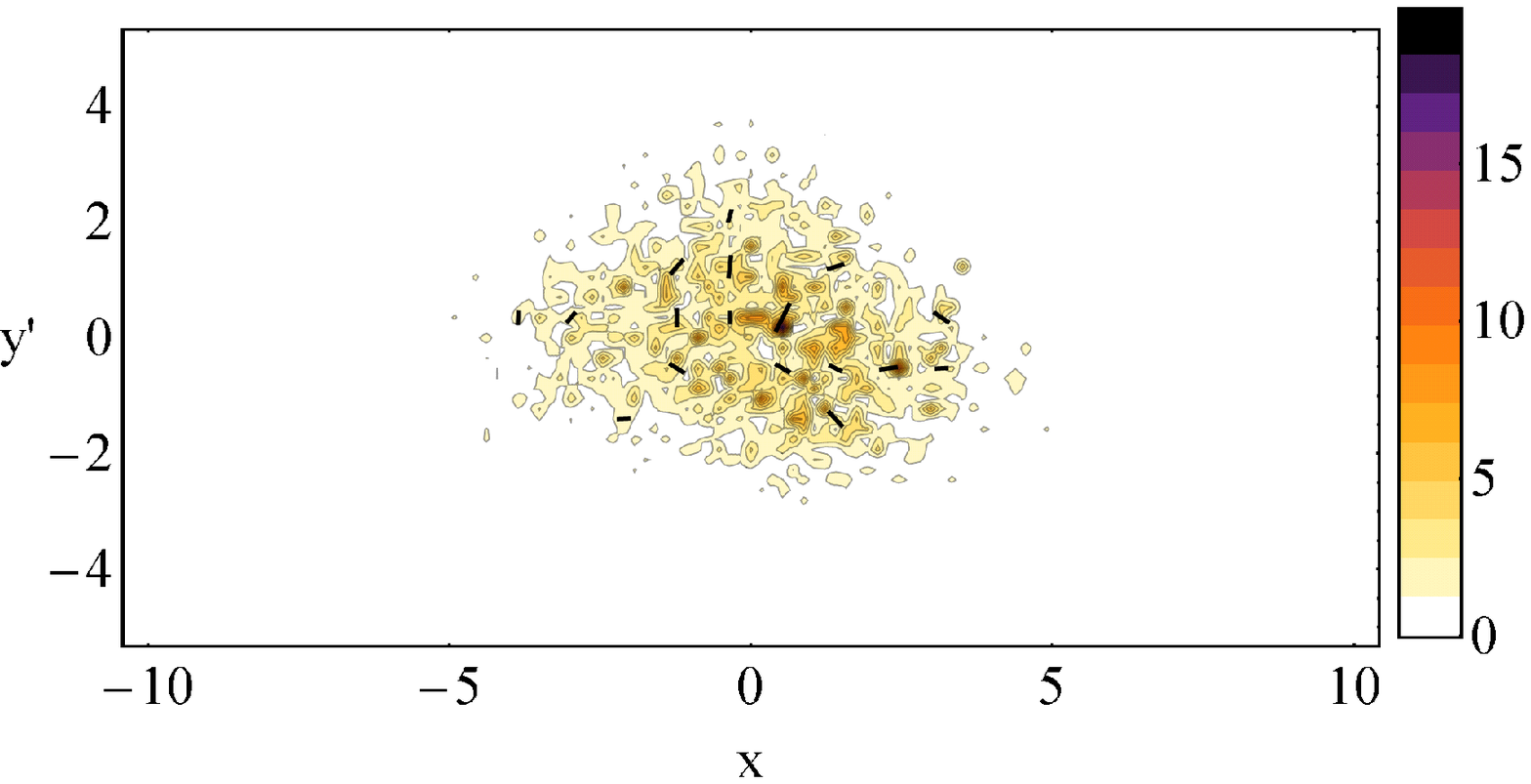}
\includegraphics[width=0.33\textwidth]{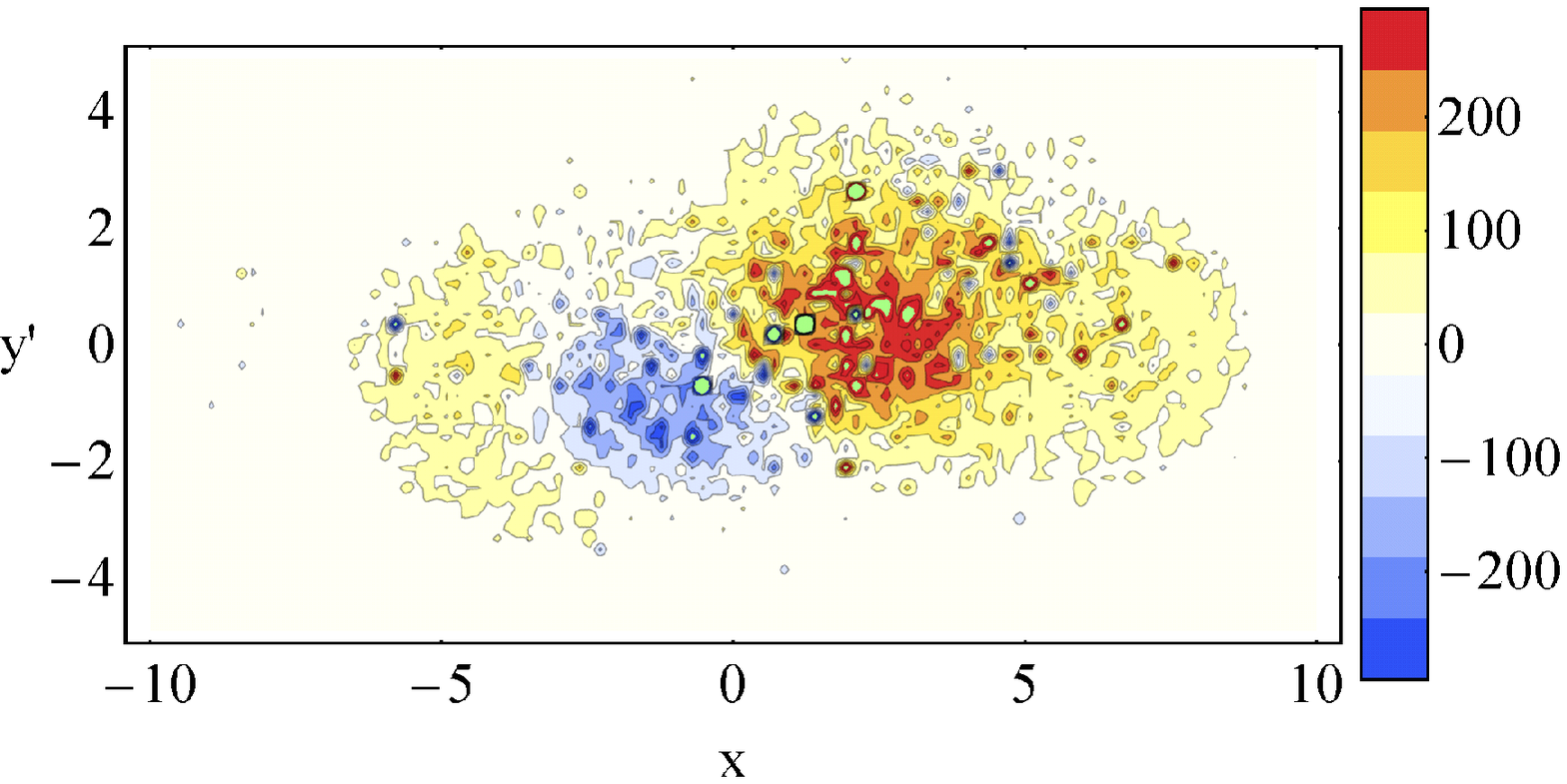}
}
\put(80,80){
\includegraphics[width=0.33\textwidth]{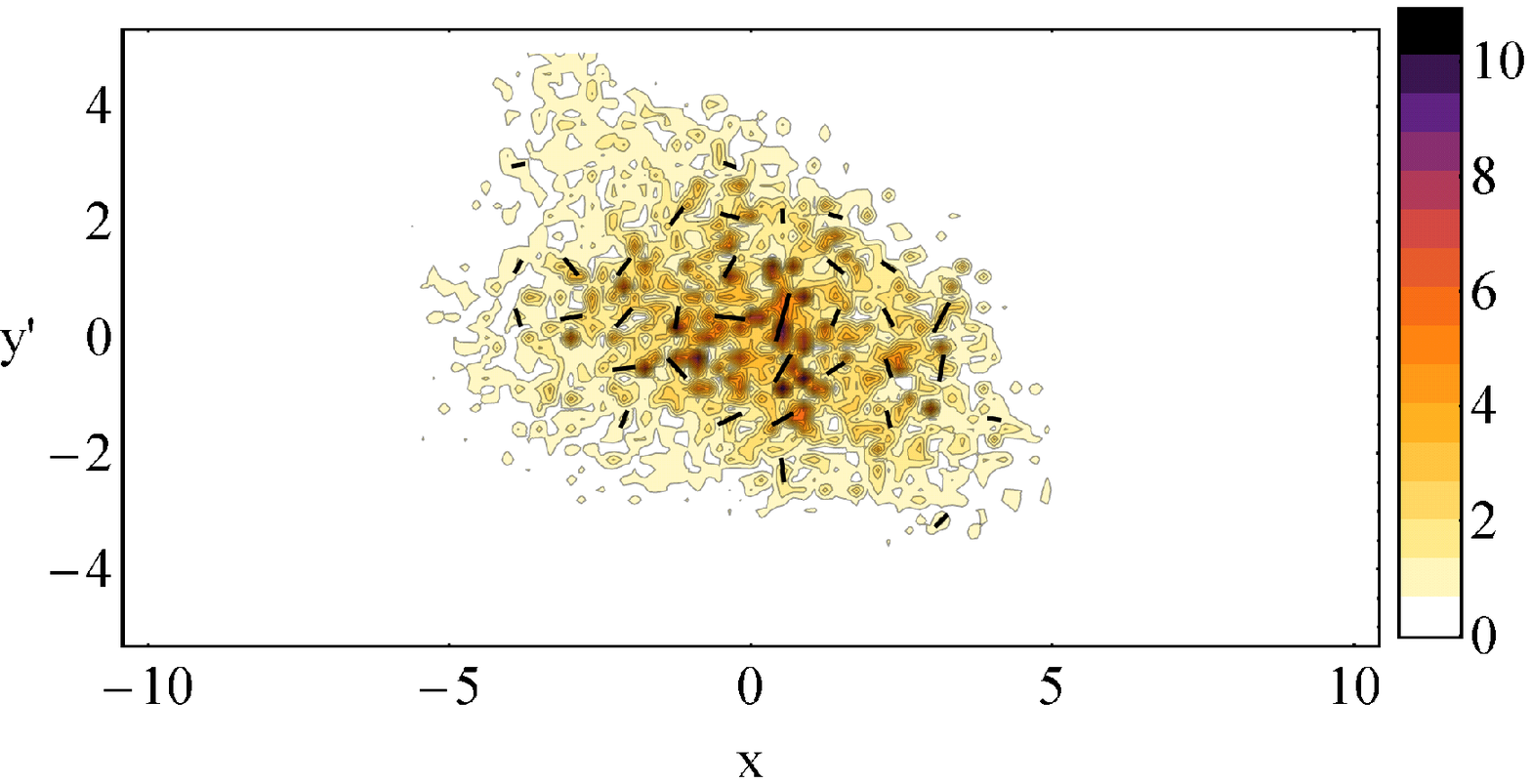}
\includegraphics[width=0.33\textwidth]{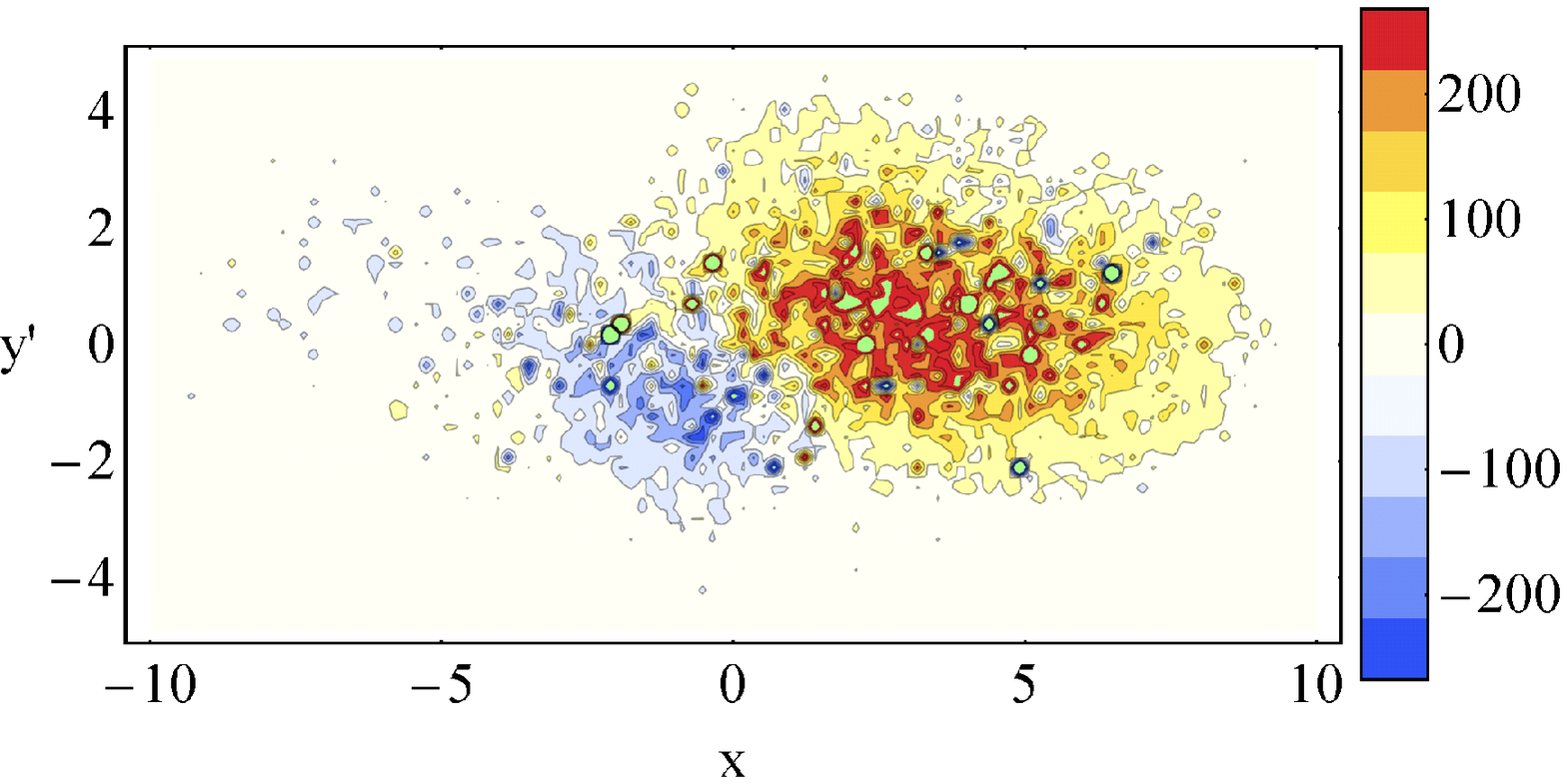}
}
\put(80,0){
\includegraphics[width=0.33\textwidth]{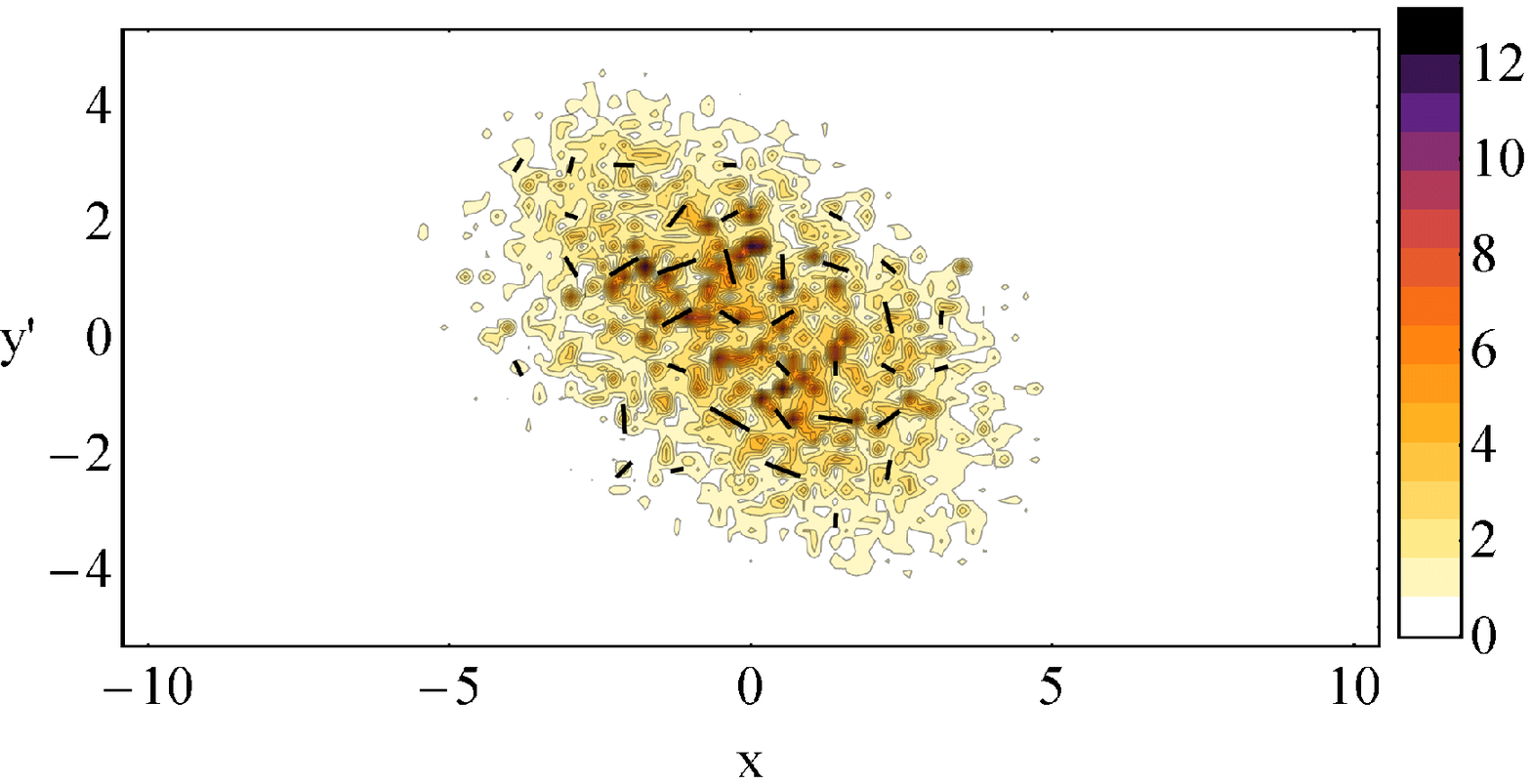}
\includegraphics[width=0.33\textwidth]{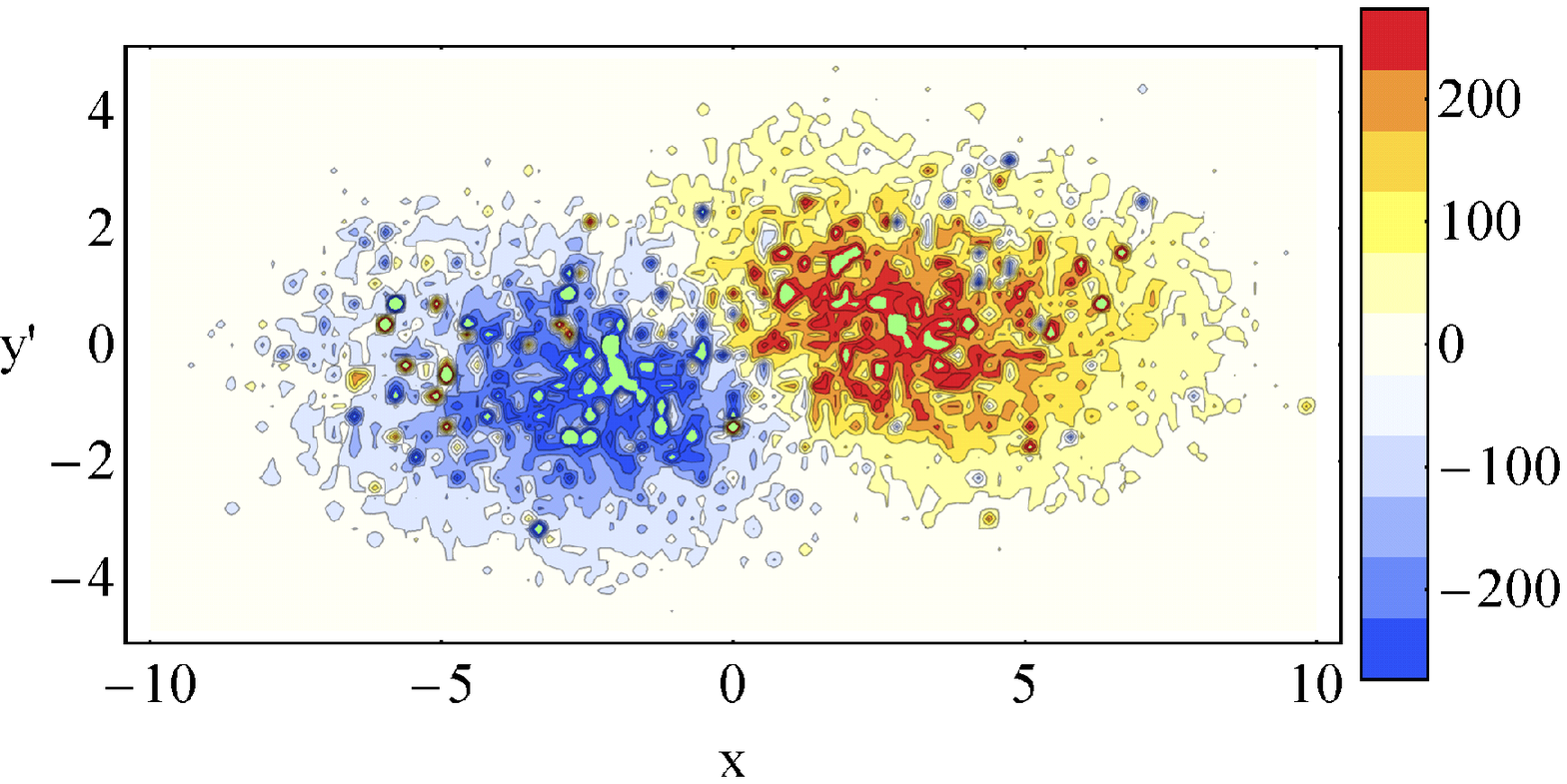}
}

\put(150,325){polarized}
\put(320,325){RM}
\put(100,310){0.8\,Gyr}
\put(100,230){2\,Gyr}
\put(100,150){3\,Gyr}
\put(100,70){13\,Gyr}
\end{picture}

\caption{The \emph{rest frame}\ simulations of
polarization (\emph{left panel}) and Faraday rotation (\emph{right
panel}) at {\bf 150\,MHz}, otherwise as in Fig.~\ref{fig:sim5ghz}. }
\label{fig:sim150mhz}
\end{figure*}

Simulations in the rest frame of the star-forming galaxy at 150\,MHz are
shown in Fig.~\ref{fig:sim150mhz}. The main difference between the 5
GHz and 150\,MHz maps is the effect of Faraday depolarization which
becomes stronger at low radio frequencies. Depolarization along the
major axis of a galaxy is stronger at large ages when the regular
field becomes stronger and more ordered. The depolarization shapes
the elongated polarized structure aligned near the minor axis of a
galaxy (Fig.~\ref{fig:sim150mhz}) which is clearly visible at
13\,Gyr after the disk formation.

We also simulate the total intensity, polarization and Faraday depth
in the rest frame of the observer at $\nu_{\rm obs}=150$\,MHz
(Fig.~\ref{fig:of150mhz}) taking into account the frequency shift of
a galaxy at a redshift $z$ (see Appendix~A for details). For more
distant galaxies the synchrotron intensity, depolarization effect,
and the regular magnetic field become weaker. The latter two effects
lead to the small ``observed'' Faraday rotations at high redshifts
(Fig.~\ref{fig:of150mhz}). The observed intensity at 150\,MHz is
emitted at high frequencies at high redshifts
(Table~\ref{tab:simipirm}).

\begin{figure*}
\centering

\begin{picture}(550,330)(0,0)
\put(80,240){
\includegraphics[width=0.33\textwidth]{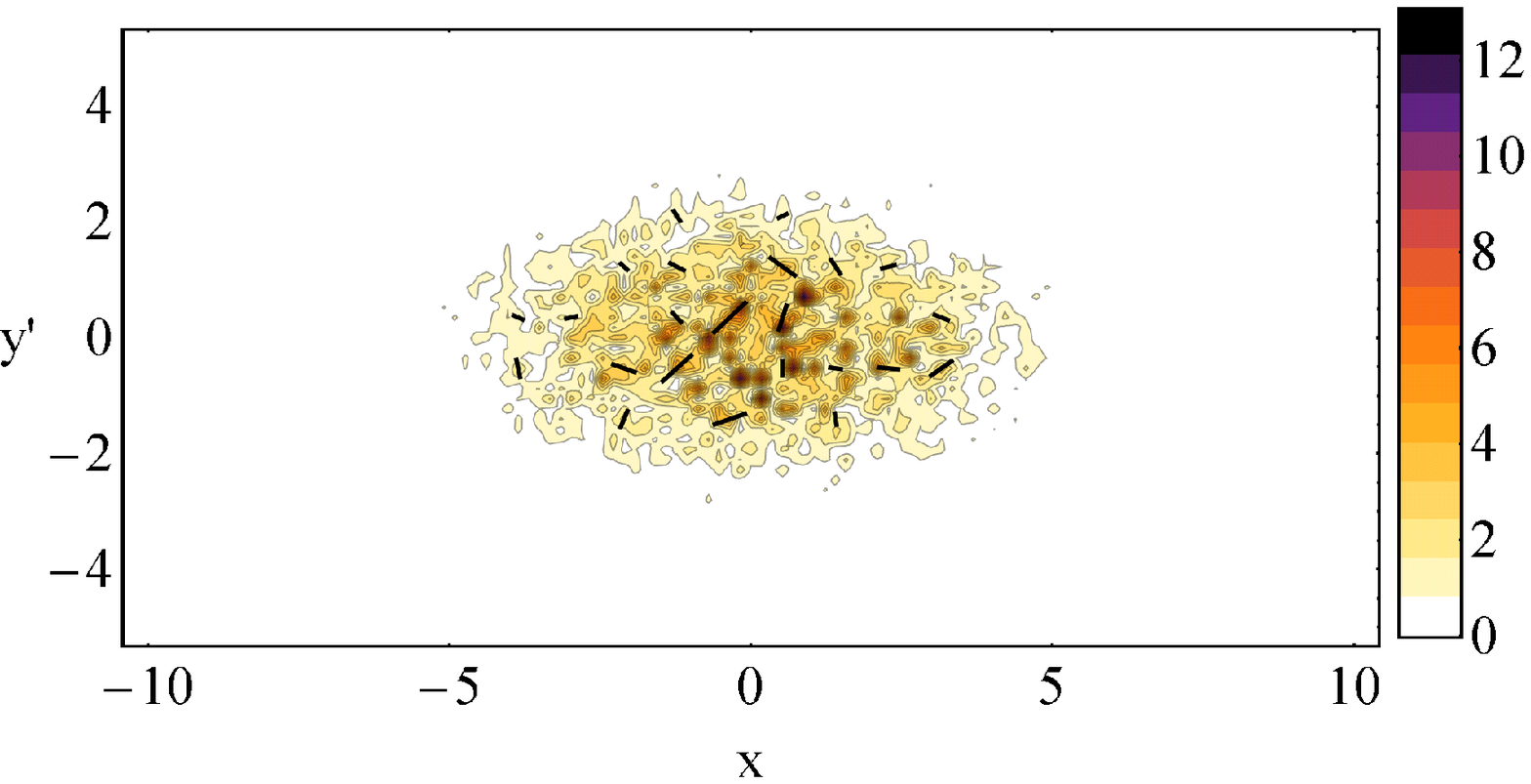}
\includegraphics[width=0.33\textwidth]{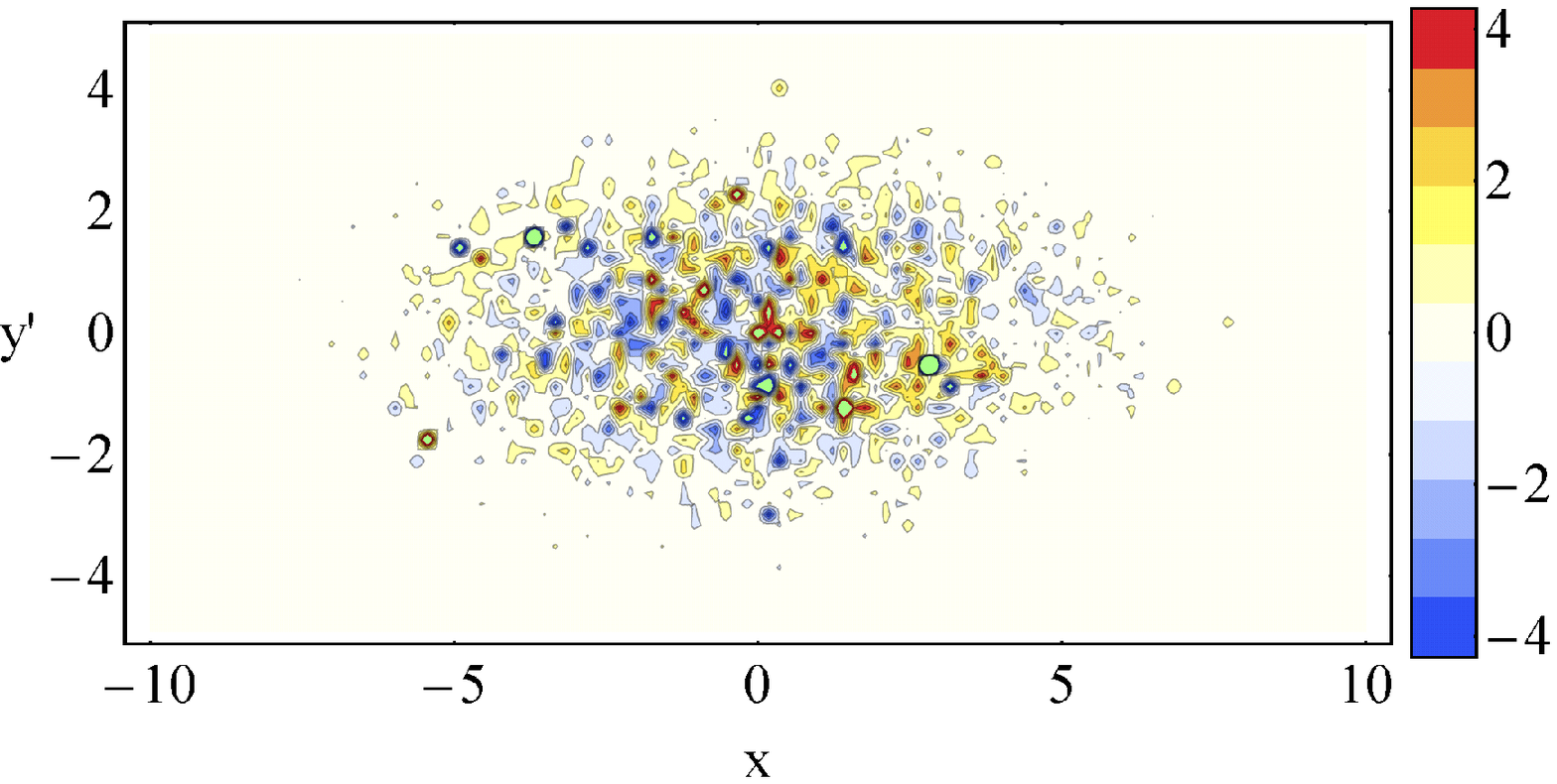}
}
\put(80,160){
\includegraphics[width=0.33\textwidth]{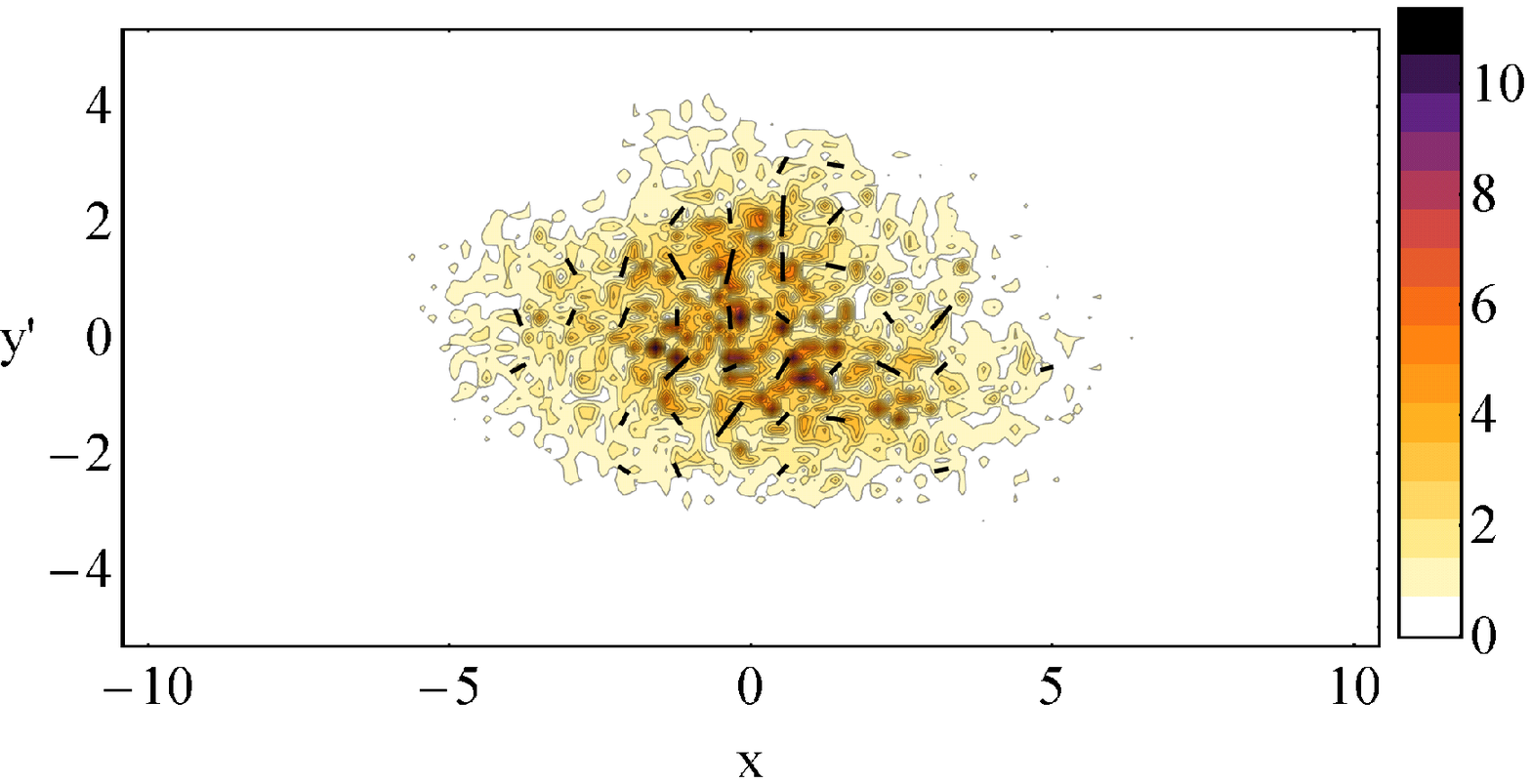}
\includegraphics[width=0.33\textwidth]{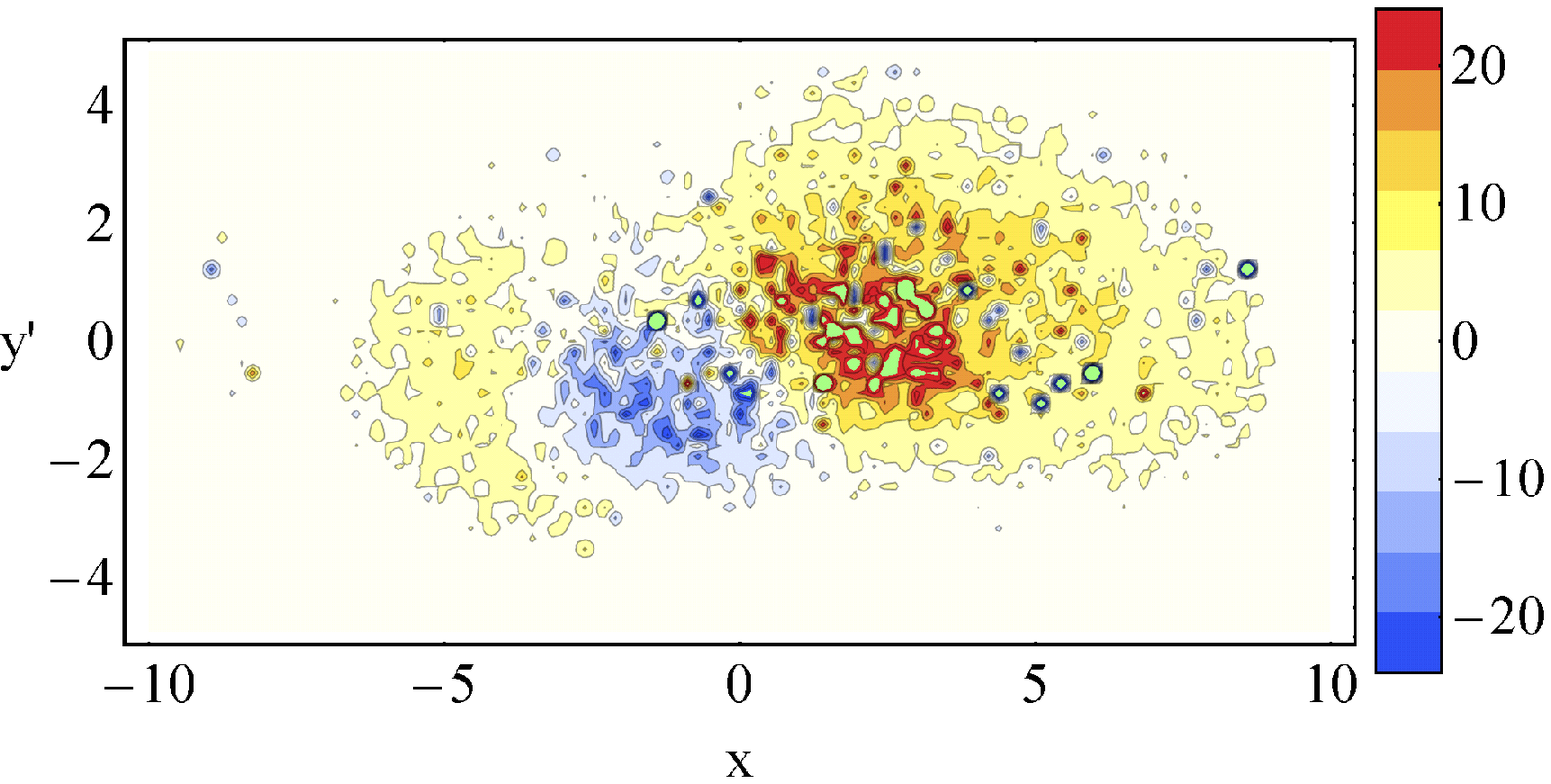}
}
\put(80,80){
\includegraphics[width=0.33\textwidth]{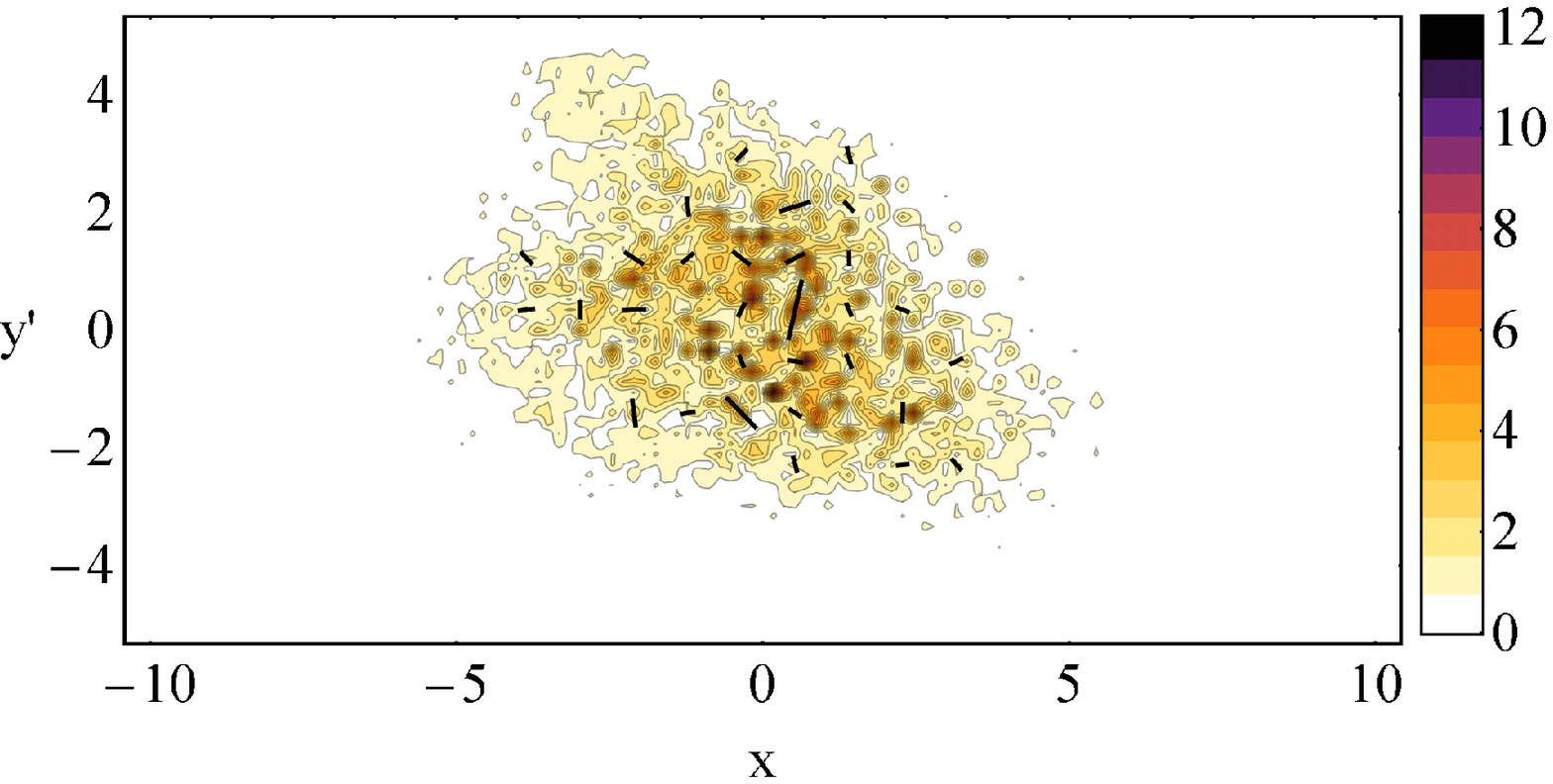}
\includegraphics[width=0.33\textwidth]{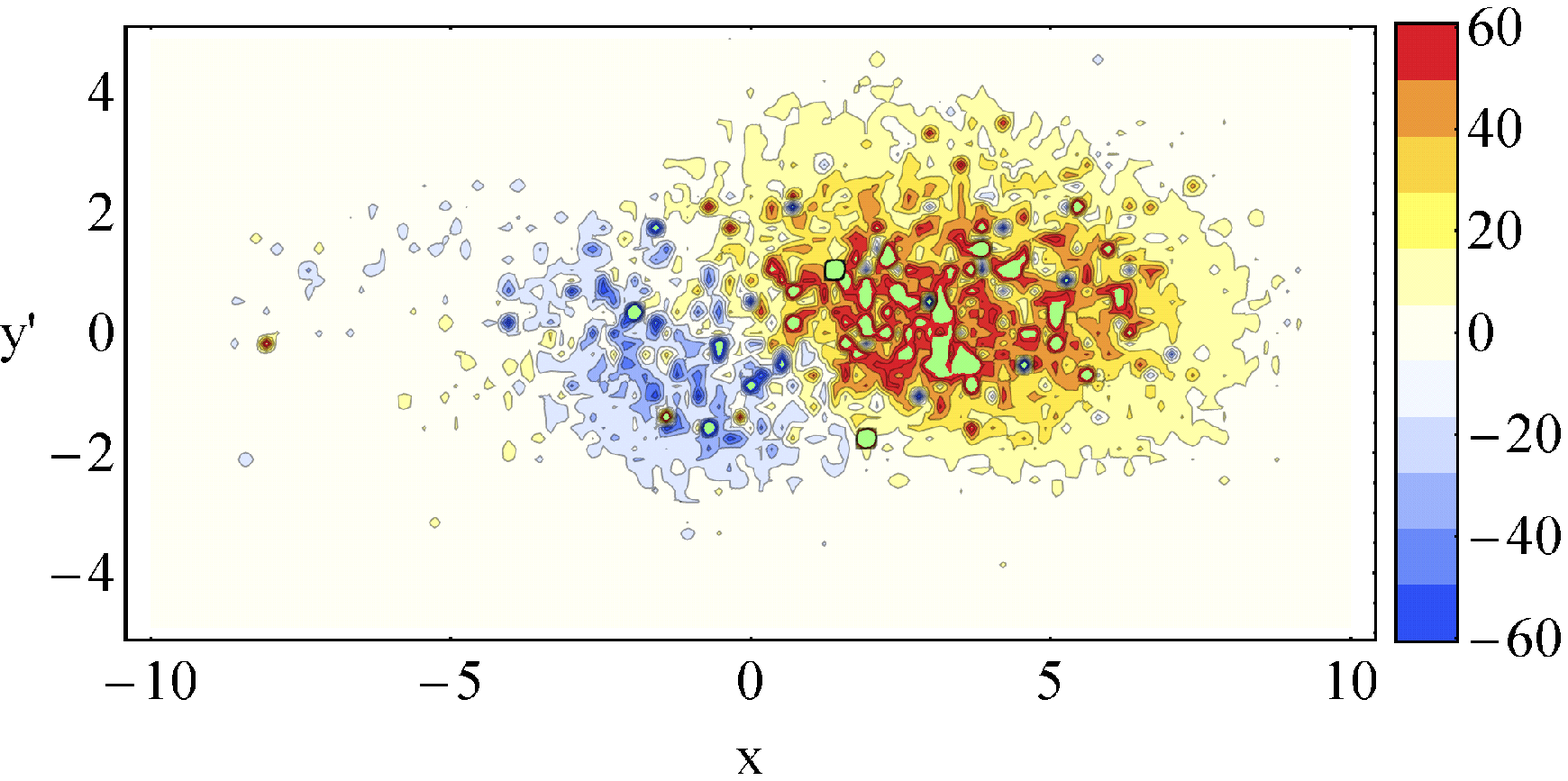}
}
\put(80,0){
\includegraphics[width=0.33\textwidth]{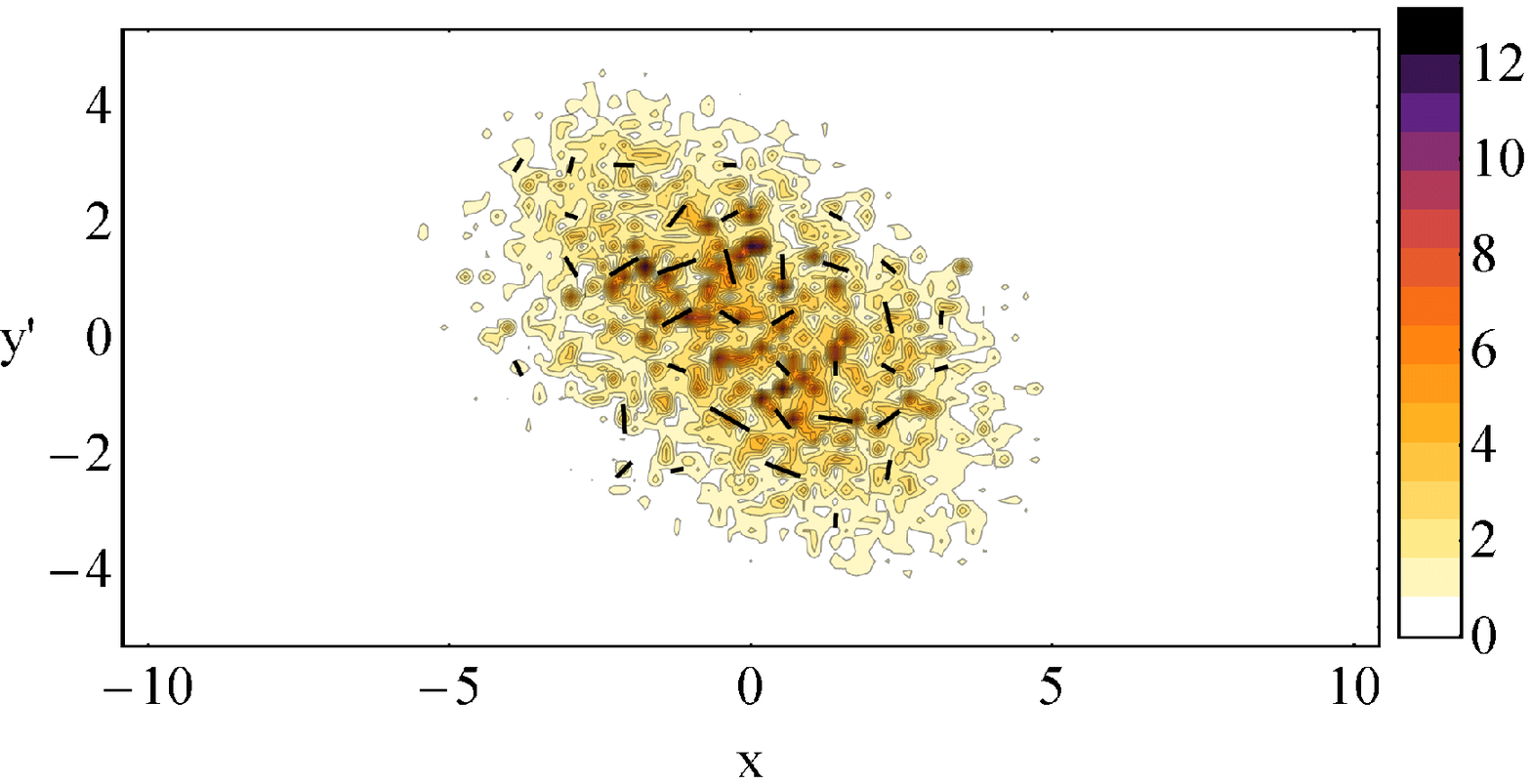}
\includegraphics[width=0.33\textwidth]{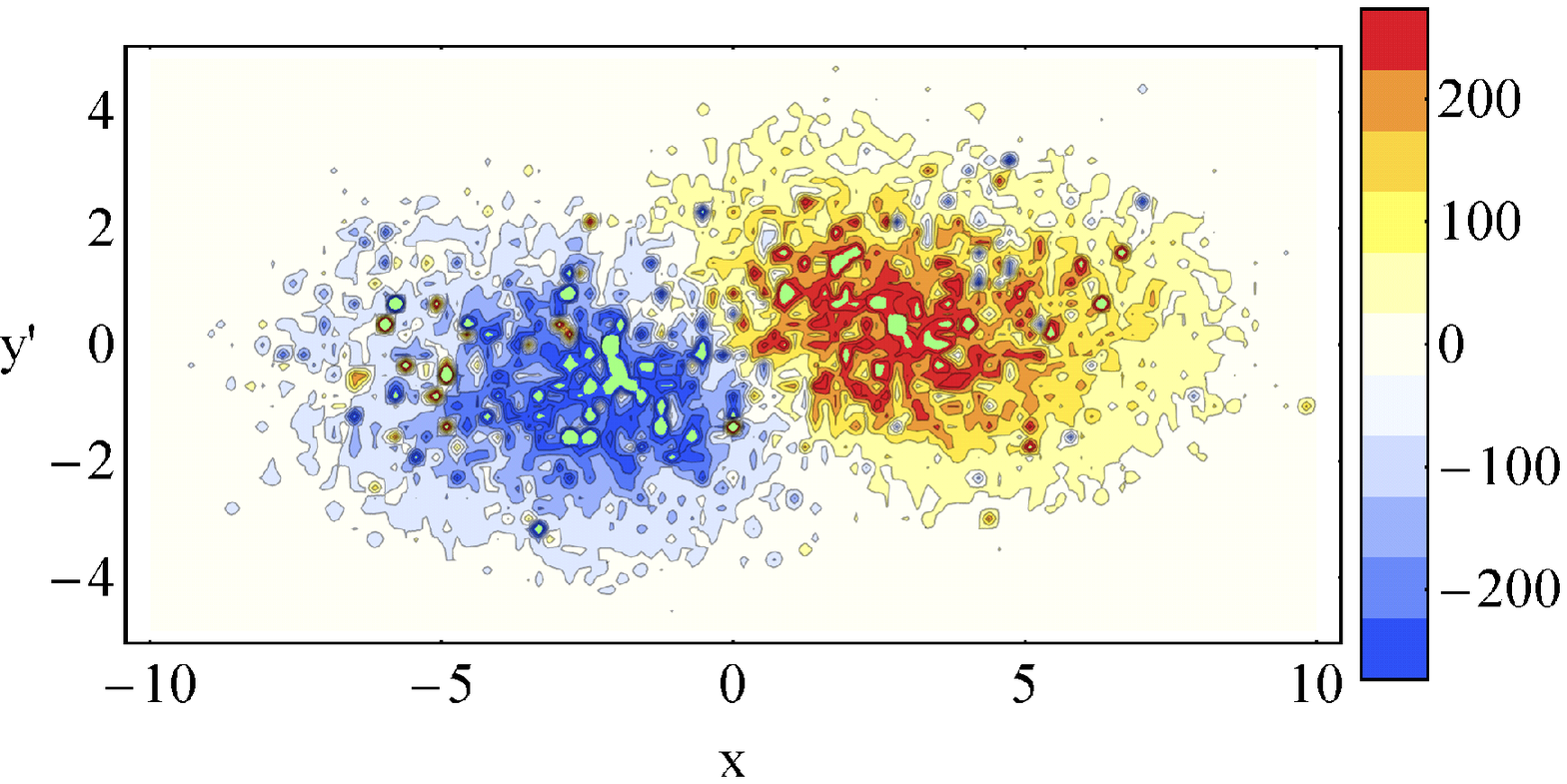}
}

\put(150,325){polarized}
\put(320,325){RM}
\put(100,310){0.8\,Gyr}
\put(100,230){2\,Gyr}
\put(100,150){3\,Gyr}
\put(100,70){13\,Gyr}
\end{picture}

\caption{Simulations of the \emph{observed}\ polarization
(\emph{left panel}) and Faraday rotation (\emph{right panel}) at
{\bf 150\,MHz}, otherwise as in Fig.~\ref{fig:sim5ghz}.}
\label{fig:of150mhz}
\end{figure*}

\section{Discussion and conclusions}
\label{sec:discussion}

The evolution of large-scale magnetic fields in galaxies has not
been investigated in detail so far. Dynamo theory predicts the
generation of large-scale coherent field patterns (``modes''), but
the timescale of this process is comparable to that of the galaxy
age (Sect.~2.2). Many galaxies are expected not to host fully
coherent fields at the present epoch, especially those which
suffered from major mergers or interactions with other galaxies.

\cite{gissinger09} presented the first global numerical
simulation of a dynamo including superbubbles generated by SNRs, but
without explicit inclusion of the alpha-effect (no mean-field assumption).
The generated spiral field is strong and of quadrupolar symmetry,
confirming the mean-field results, and it resembles the "spotty"
field injection.

Kinematical simulations of the field evolution in spiral galaxies
without feedback to the gas flow, but excluding dynamo action
\citep{linden98} revealed large-scale field structures which are
partly oriented along the spiral arms, but also features of regular
fields in the interarm regions which rapidly change their
structure within one rotation of the galaxy and hence do not lead to
a stable pattern. Presently it is unknown whether feedback of the
field onto the gas flow may stabilize the field pattern.

The first global galactic-scale MHD simulation of a CR-driven dynamo
with 100\,pc resolution was performed by \cite{hanasz09} in which
the cosmic rays are produced in randomly occurring supernova
remnants. The timescales for amplification and ordering of the
magnetic filed in their model are consistent with the dynamo
timescales estimtated by \cite{arshakian09}. This gives some
confidence that the dynamo model correctly describes the overall
properties of field evolution, especially as the simulations by
\cite{hanasz09} also show ``spotty'' magnetic structures during the
evolution from randomly spread seed fields to large-scale spiral
structures. Even the ratio $h/R$ regulates itself in the CR--MHD
simulations to a value near 0.1 which we assumed in the dynamo
models (Hanasz, private communication).

Spiral arms and self gravity of the gas need to be included in
future improvements of the model. The inclusion of shock fronts from
supernova remnants will need higher spatial resolution. Furthermore,
the effects of the SFR on the strength of the
amplification and saturation of the regular magnetic field has still
to be included into the model. As the SFR decreases with decreasing
redshift and depends on merger rate, this may also influence the
evolution of the magnetic field in evolving galaxies.

\emph{Predictions of the model.} Comparison of our predictions with observations has to be
restricted to spiral galaxies with thin disks and to high-frequency
observations. Also, comparison is still restricted to nearby
galaxies which can be resolved with the limited sensitivity of
present-day telescopes. Among the existing observations, we could
find spiral galaxies with well-devel\-oped large-scale spiral
(axisymmetric) magnetic field structures, like in M\,31
\citep{beck82, berkh03}, IC\,342 \citep{krause89} and NGC\,6946
\citep{beck07}, as visible in polarized intensity and Faraday
rotation similar to the last row in Fig.~\ref{fig:sim5ghz}. All
three galaxies are large with an exponential scale radius of about
5\,kpc. Hence, according to \cite{arshakian09} it should have taken
about 9\,Gyr for these galaxies to coherently order the regular
field structure over the entire disk (twice scale radius). This
indicates that they did not suffer any major merger during this
period. Indeed, NGC\,6946 and IC\,342 are isolated galaxies without
any known companion and without any signs of major tidal
distortions.

Observations of M\,51 showed, however, that even if a large-scale
spiral magnetic field is present in the galaxy's disk and apparent
in polarized intensity, it is hidden in the map of Faraday rotation
\citep{fletcher10} and looks much more patchy than that in row 3 of
Fig.~\ref{fig:sim5ghz}. This may be due to an additional magnetic
field component, probably an isotropic turbulent magnetic field with
scales of the longer axis between 400\,pc to 1\,kpc. Such a field
has been found in M\,51 \citep{fletcher10} and in the barred galaxy
NGC~1097 \citep{beck05b}, but was not included into our simulations
shown in Fig.~\ref{fig:sim5ghz} and Fig.~\ref{fig:sim150mhz}. This
component may be produced by strong shearing gas motions or
compressions. It may also form loops perpendicular to the galaxy's
disk due to the Parker instability.

A prediction of our model is that interacting and merging
galaxies should reveal complicated field patterns. The nearby and
best studies case of strongly interacting/mer\-ging galaxies is the
Antennae galaxy pair NGC\,4038/39 observed in total and
polarized radio emission by \cite{chyzy04}. A huge polarized ridge
in the northeast is probably the result of shearing flows. Another
highly polarized region between the two galaxies marks the location
of stron\-gly compressed fields. The spiral field pattern of the
galaxy in the northwest is still clearly visible. This galaxy is
still in the phase before total field disruption and cannot be
compared with our simulations.

A survey of polarized emission from galaxies in the Vir\-go cluster by
\cite{wez07} and \cite{wez10} revealed signs of compression or shear
in almost all galaxies, but the Faraday rotation data are still of
insufficient quality to detect any delay in the formation of
coherent fields.

Another prediction from our model, that small galaxies build up
their large-scale field faster than large galaxies, also cannot be
observationally confirmed with the existing data from nearby spiral
galaxies. We need a sample of galaxies without massive density waves
and without bars where the dynamo-generated field dominates over
anisotropic fields. Furthermore, galaxies with strong signs of tidal
interactions have to be excluded. A comparison with observations of
dwarf and irregular galaxies fails because they do not fulfill the
``thin-disk criteria'' ($h/R < 0.1$) discussed in Sect.~\ref{ssec:assumptions}.

Forthcoming observations of a large sample of spiral galaxies with
high resolution and high sensitivity and of distant galaxies in
early stages of the evolution with the SKA and its precursors
should be able to test our scenario \citep{beck09}.

\emph{Limitations of the model.} Lacking a realistic MHD model of evolving galaxies, we tried in this
paper to describe the evolution of large-scale fields by simple
interpolation between the stages about which we believe to have some
knowledge: the epoch of field seeding and the present epoch.
This approach provided templates for radio maps in total and
polarized intensity and in Faraday rotation measures, as part of the
simulations for the SKA Design Studies (SKADS).

In Arshakian et al. (2009), we assumed that the ``quasi-spherical''
mean-field dynamo amplified the regular field and increased the
coherence scale in radial direction, while the ``disk'' mean-field
dynamo was effective in thin-disk galaxies. We note that the
simulations of the ``disk'' mean-field dynamo described in the
previous section cannot be performed for dwarf and irregular
galaxies.

Another limitation of simulations comes from the fact that the
strength of the cosmic microwave background energy density increases
with increasing redshift, so that the synchrotron emission of
star-forming galaxies will be quick\-ly suppressed by inverse Compton
losses off the cosmic microwave background at redshifts $z>3$
\citep{murphy09}. In this paper, we do not account for this effect,
but our simulations should be realistic for star-forming galaxies up
to $z \approx 3$.

High SFR causes high velocity turbulence of the ionized gas in
starburst galaxies, e.g. by a major merger of gas-rich galaxies. If
SFR is higher than or equal to 20\,$M_{\sun}$\,yr$^{-1}$, the
mean-field dynamo will be suppressed \citep{arshakian09}. Hence, our
simulations are valid only for disk galaxies with ${\rm SFR}\la
20$\,$M_{\sun}$\,yr$^{-1}$.

In reality, the magnetic field evolution in first
galaxies can be affected by various effects which are not accounted in our simple model. In particular, galactic winds \citep[e.g.][]{moss10}, turbulent pumping
 \citep[e.g.][]{brandenburg95}, and the multiscale nature of the interstellar
medium \citep{mantere10} can substantially
affect the dynamo and result in a phenomenology which observably
differs from predictions of our scenario. Future
observations with SKA and its forthcoming pathfinders will be crucial to detect 
the difference with our simplified model and prove the importance of additional
effects.

Furthermore, our simulations assume that the field structure in the
halo is the same as in the disk and do not take into account the
X-shaped halo fields observed around edge-on galaxies
\citep{krause09}. When observing mildly inclined galaxies at low
frequencies, the polarized emission from the disk is mostly
depolarized and the halo may dominate. In this case, our results for
polarized and Faraday rotation images shown in
Fig.~\ref{fig:sim150mhz} may change significantly. This can be
tested by sensitive observations of edge-on galaxies at sufficiently
high frequencies to avoid Faraday depolarization. \\

In summary, we link the SFR and amplitudes of the regular and
turbulent fields in disk galaxies, further develop an evolutionary
model for magnetic fields in isolated star-forming disk galaxies and
performed modeling of the evolving galaxy. The
amplitude of the initial ``spotty" structure  of regular fields
(generated at the epoch of disk formation) is amplified by means of
mean-field dynamo to the equipartition level in a few Gyr and
remains at this level up to the present time. The ordering and
azimuthal scales of initial magnetic spots increase with the age of
a galaxy stretching the initial field in both directions. For the
proposed model, we simulate the total intensity, polarization and
Faraday rotation for the MW-type galaxy at different frequencies
(from 150\,MHz to 5\,GHz) observable with the SKA and its
precursors. A number of predictions such as the patchy structure and
field reversals of regular fields, patchy Faraday rotations in
galaxies younger than few Gyrs, smaller ordering scale in young
galaxies, polarization patterns at 5\,GHz and asymmetric structure
at 150\,MHz in older galaxies because of stronger depolarization,
and the complicated field patterns in interacting and merging (minor
and major) galaxies can be tested with the future radio telescopes.

\acknowledgements We are grateful to Anvar Shukurov for fruitful
discussions. We thank the anonymous referee for valuable comments.
Numerical simulations were partially performed on the
supercomputer ``Tchebyshoff'' at the Computing Center of Moscow
University. TGA acknowledges the grant 566960 by DFG-SPP project.
RS  acknowledges the  grant YD-4471.2011.1 by the Council of the President of the
Russian Federation and the RFBR grant
11-01-96031-ural. This work is supported by the European Community Framework Programme
6, Square Kilometre Array Design Study (SKADS),  and the DFG-RFBR
project under grant 08-02-92881.

%\newpage

\begin{appendix}
\section{Equations for a total and polarized intensities, and Faraday rotation}
The polarized signal passes a distance $w'$ in the direction of the
line-of-sight through a magneto-ionic medium of the galaxy inclined at an angle $i$. Then the Faraday rotation (in units of rad m$^{-2}$), polarization angle, and total intensity in the coordinate system inclined at an angle $i$ $(x, y'=y\cos i, w'=w\cos i)$ are
\begin{equation}
RM(x,y',w') = 810 \int_{-\infty}^{w'}   B_\parallel(\vec r)n_{\rm{e}}(\vec r) ds,
\end{equation}

\begin{eqnarray}
  \lefteqn{ \chi(x,y',w',\lambda,z) =  \chi_0(x,y',w') + } \nonumber \\
  & & \left(\frac{\lambda}{1+z} \right)^2 RM(x,y',w'),
\end{eqnarray}
and
\begin{equation}
I(x,y',\lambda,z) = \left(\frac{\lambda}{1+z} \right)^{\alpha}\int_{-\infty}^{\infty}  n_c(\vec r)  |B_{\bot}(\vec r)|^{1+\alpha}  ds,
\end{equation}
where $B$ is the magnetic field strength measured in $\mu$G, $n_{\rm{e}}$ in cm$^{-3}$, $ds$ and $w'$ in kpc, $\chi_0$ is the intrinsic magnetic field orientation (from the
magnetic field model), $\lambda$ is the observed wavelength (in meters), $z$ is the redshift of a galaxy, $\alpha$ is the synchrotron spectral index ($\alpha=0.8$). The equations for Stokes parameters, $Q$ and $U$, are given by
\begin{eqnarray}
  \lefteqn{ Q(x,y',\lambda,z) = } \nonumber \\
  & & p_i\left(\frac{\lambda}{1+z} \right)^{\alpha} \int_{-\infty}^{+\infty} n_c(\vec r) |B_{\bot}(\vec r)|^{1+\alpha} \cos(2\chi) ds,
\end{eqnarray}
\begin{eqnarray}
  \lefteqn{ U(x,y',\lambda,z) = } \nonumber \\
  & & p_i\left(\frac{\lambda}{1+z} \right)^{\alpha} \int_{-\infty}^{+\infty} n_c(\vec r) |B_{\bot}(\vec r)|^{1+\alpha} \sin(2\chi) ds,
\end{eqnarray}
where $p_i=0.73$ is the intrinsic maximum polarization for
synchrotron emission (for $\alpha=0.8$) in regular magnetic fields.
Then, the ``observed'' polarization intensity
\begin{equation}
PI(x,y',\lambda) = \sqrt{Q(x,y',\lambda)^2 + U(x,y',\lambda)^2},
\end{equation}

\begin{equation}
\chi_{\rm obs}(x,y',\lambda) = \frac{1}{2} \arctan\frac{U(x,y',\lambda)}{Q(x,y',\lambda)},
\end{equation}
and the ``observed'' RM is
\begin{equation}
RM_{\rm obs}(x,y',\lambda) = \frac{\partial \chi(x,y',\lambda^2)}{\partial\lambda^2}.
\end{equation}

\end{appendix}

\end{document}